\documentclass[11pt]{article}
\usepackage[dvips]{color}                 
\usepackage{graphicx}                     
\usepackage{amssymb}
\usepackage{amsmath}
\usepackage{xspace}   
\usepackage[latin1]{inputenc}
\usepackage[english]{babel}
\usepackage{amsfonts}
\usepackage{epsfig}
\usepackage{changebar}
\usepackage{slashed}

\renewcommand{\today}{\ifcase\day\or 1st\or 2nd\or 3rd\or 4th\or 5th\or 6th\or
 7th\or 8th\or 9th\or 10th\or 11th\or 12th\or 13th\or 14th\or 15th\or 
 16th\or 17th\or 18th\or 19th\or 20th\or 21st\or 22nd\or 23rd\or 24th\or
 25th\or 26th\or 27th\or 28th\or 29th\or 30th\or 
 31st\fi~\ifcase\month\or January\or February\or March\or April\or
 May\or June\or July\or August\or September\or October\or November\or
 December\fi \space \number\year}   
\newcommand{\mytitle}[1]{
                         \begin{center}
                           \LARGE{\textbf{#1}}
                         \end{center}}
\newcommand{\myauthor}[1]{\textbf{#1}}
\newcommand{\myaddress}[1]{\textit{#1}}
\newcommand{\mypreprint}[1]{\begin{flushright} #1 \end{flushright}}
\begin{document}

\def\sl#1{\slash{\hspace{-0.2 truecm}#1}}
\def\g{g_{A}}
\def\F{F_{\pi}}
\def\m{m_\pi}
\def\M{M_0}
\def\I{I_{11}}
\def\Mn{M_N}
\def\tM{\tilde{M}}

\begin{titlepage}
\mypreprint{
TUM-T39-07-04}

\vspace*{0.5cm} 
\mytitle{Chiral Perturbation Theory and the first moments of the Generalized Parton Distributions in a 
Nucleon\footnote{Work supported in part by BMBF and EU-I3HP}}
  \vspace*{0.3cm}

\begin{center}
 \myauthor{Marina Dorati$^{a,b}$},
\myauthor{Tobias A. Gail$^b$}
and
  \myauthor{Thomas R. Hemmert$^{b}$} 

  \vspace*{0.5cm}
\myaddress{$^a$ Dipartimento di Fisica Nucleare e Teorica \\
      Universita' degli Studi di Pavia and INFN, Pavia, Italy}\\[2ex]
  \myaddress{$^b$
    Physik-Department, Theoretische Physik T39  \\
    TU M{\"u}nchen, D-85747 Garching, Germany}\\[2ex]

  \vspace*{0.2cm}
\end{center}

\vspace*{0.5cm}

\begin{abstract}
\noindent
We discuss the first moments of the parity-even Generalized Parton Distributions (GPDs) in a nucleon, corresponding to six (generalized) vector form 
factors.
We evaluate these fundamental properties of baryon structure at low energies, utilizing the methods of covariant 
Chiral Perturbation Theory in the baryon sector (BChPT). Our analysis
is performed at leading-one-loop order in BChPT, predicting both the momentum and the quark-mass dependence for the three 
(generalized) isovector and (generalized) isoscalar form factors, which are currently under investigation in lattice QCD analyses of
baryon structure. We also study the 
limit of vanishing four-momentum transfer where the GPD-moments reduce to
the well known moments of Parton Distribution Functions (PDFs). For the isovector moment $\langle x\rangle_{u-d}$ our BChPT
calculation predicts a new mechanism for chiral curvature, connecting the high values for this moment typically found in lattice QCD studies for
 large quark masses with the smaller value known from phenomenology.  Likewise, we analyze the quark-mass dependence of the 
isoscalar moments
in the forward limit and extract the contribution of quarks to the total spin of the nucleon. We close with a first glance at the momentum 
dependence of the isoscalar C-form factor of the nucleon.
\end{abstract}
\end{titlepage}
\section{Introduction}
\setcounter{footnote}{0}

Understanding the structure of the nucleon arising from the underlying dynamics of quarks and gluons---governed by the theory of 
Quantum-Chromo-Dynamics (QCD)---is still one of the major challenges 
in nuclear physics. About a decade ago the concept of Generalized Parton Distributions (GPDs) has emerged among theorists, constituting
a universal framework bringing a host of seemingly disparate nucleon structure observables like form factors, moments of parton distribution
functions, etc. under one theoretical roof. For reviews of this very active field of research we refer to \cite{reviews}.
 
Working in twist-2 approximation, the parity-even part of the structure of the nucleon  
is encoded via two Generalized Parton Distribution functions
$\textit{H}^{q}(x,\xi,t)$ and $\textit{E}^{q}(x,\xi,t)$. For a process where the incoming
(outgoing) nucleon carries the four-momentum $p_1^\mu\,(p_2^\mu)$ we define two new momentum variables 
\begin{equation}
\Delta^\mu=p_2^\mu-p_1^\mu;\quad\bar{p}=(p_1^\mu+p_2^\mu)/2.
\end{equation}
The GPD variable $x$ then corresponds to
Bjorken-$x$, which in the parton model of the nucleon can be interpreted as the fraction of the total momentum of the nucleon 
carried by the probed quark $q$. $t\equiv \Delta^2$ denotes the total four-momentum transfer (squared) to the nucleon, whereas  the 
``skewdness'' variable $\xi=-n\cdot\Delta/2$ with $n\cdot \bar{p}=1$ interpolates between the $t$- and the $x$-dependence of the GPDs. 
Throughout this work we focus on the non-perturbative regime of QCD: $-t<<1$ GeV$^2$, i.e. the realm of strongly 
interacting matter in our solar system making up humans, planets, etc.. We are therefore utilizing the methods of Chiral Effective Field Theory
(ChEFT) for our analysis \cite{Chreview}.

The 3-dimensional parameter space of GPDs is vast and rich in information about nucleon structure. The experimental program of
their determination is only at the beginning at laboratories like CERN, Desy, JLAB, \ldots \cite{reviews}. However, moments of GPDs can be interpreted
much easier and are connected to well-established hadron structure observables. E.g. the zero-th order (Mellin-) moments in the variable $x$ 
correspond
to the contribution of quark $q$ to the well known Dirac and Pauli form factors $F_1(t),\,F_2(t)$ of the nucleon: 
\begin{eqnarray}
\int_{-1}^{1}dx\,x^0\,\textit{H}^{q}(x,\xi,t)&=&F_{1}^{q}(t),\\
\int_{-1}^{1}dx\,x^0\,\textit{E}^{q}(x,\xi,t)&=&F_{2}^{q}(t). \label{f2}
\end{eqnarray}
For the case of 2 light flavors the isoscalar and isovector Dirac and Pauli form factors of the nucleon have been studied at low values of $t$ 
at the one-loop level in 
Chiral Effective Field Theory, both in non-relativistic \cite{BFHM} and in covariant \cite{GSS,Kubis} schemes. The chiral extrapolation of these
form factors for lattice QCD data with the help of ChEFT\footnote{In a field theory like ChEFT the result obtained at a given order in the 
perturbative expansion is independent of the the choice of regulator applied to the UV-limit of the diagrams. (For an explicit demonstration 
we refer to the comparison between cutoff- and dimensional-regularization for the mass of the nucleon in HBChPT in ref.\cite{NRmass}). 
Nevertheless, forms of UV-regulators for HBChPT loop diagrams have been proposed in the literature, which effectively amount to modeling 
short-distance contributions
of higher orders. The resulting chiral extrapolation functions have then been utilized to extend the limit of applicability of such one loop 
calculations
towards large momentum transfer $|t|\sim 1$ GeV$^2$ and towards the heavy-quark limit. Here we do not advocate such an approach, but
for recent work in this direction we refer to 
ref.\cite{Ross}.} has been discussed in refs.\cite{HW,QCDSFff,Cyprus,Marc}, whereas the status 
of the experimental situation is reviewed in ref.\cite{kees}.

In this work we want to focus on the {\em first} moments in $x$ of these nucleon GPDs
\begin{eqnarray}
\int_{-1}^{1}dx\,x\,\textit{H}^{q}(x,\xi,t)&=&A_{2,0}^{q}(t)+(-2\xi)^{2}\textit{C}_{2,0}^{q}(t), \label{momentsA}\\
\int_{-1}^{1}dx\,x\,\textit{E}^{q}(x,\xi,t)&=&B_{2,0}^{q}(t)-(-2\xi)^{2}\textit{C}_{2,0}^{q}(t), \label{momentsB}
\end{eqnarray}
where one encounters {\em three generalized form factors} $A_{2,0}^q(t),\,B_{2,0}^q(t),\,C_{2,0}^q(t)$ of the nucleon for each quark flavor 
$q$. For the case of 2 light flavours the {\em generalized isoscalar and isovector form factors} have been analyzed in a series of papers 
at leading-one-loop order in the non-relativistic framework of HBChPT, starting with the pioneering analyses of Chen and Ji as well as Belitsky and Ji 
\cite{CJ2,BJ,ACK,DM}. In this work we want to provide the first analysis of these generalized form factors utilizing the methods of
covariant Baryon 
Chiral Perturbation Theory (BChPT) for 2 light flavors, pioneered in ref.\cite{GSS}. Utilizing a variant of Infrared Regularization \cite{BL} 
for the loop
diagrams, our BChPT formalism is constructed in such a way that we {\em exactly} reproduce\footnote{We note that this property is not shared by all
regularization schemes proposed for covariant BChPT during the past few years \cite{GH}.} the corresponding HBChPT result of 
the same chiral order in the limit of small pion masses\footnote{Technically speaking, for such a comparison one has to expand the results 
of loop diagrams in BCHPT in terms of $1/(16\pi^2F_\pi^2 M_0^n)$, where $n$ is determined by the order of the HBChPT calculation and
$F_\pi$ denotes the pion decay constant.}. 

We note that (at $t=0$) a covariant BChPT calculation differs from a non-relativistic one---provided both are 
performed at the same chiral order $D$---by an infinite series of terms $\sim (m_\pi/M_0)^n$, where $M_0$ denotes the mass of the nucleon
 (in the chiral limit), 
estimated to be around 890 MeV \cite{Mass}. Such terms quickly become relevant once the pion
mass $m_\pi$ takes on values larger than 140 MeV, as it typically occurs in present-day lattice QCD simulations of (generalized) form factors.
Aside from this resummation property in $(1/M_0)^n$, we note 
that the power-counting analysis determining possible operators and allowed topologies for loop diagrams 
at a particular chiral order (see section \ref{counting}) is {\em identical} between covariant and non-relativistic frameworks. 
Both schemes organize a perturbative calculation as a power series 
in $(p/\Lambda_\chi)^D$, where $p$ corresponds either to a small 3-momentum or to the mass of the pion, whereas 
$\Lambda_\chi\approx 1.2$ GeV denotes the scale of chiral symmetry breaking. Finally, we would like to note that the first moments of 
nucleon GPDs have also been studied in constituent quark models ({\it e.g.} see ref.\cite{Pavia})  and chiral quark soliton models ({\it e.g.} see 
ref.\cite{Bochum}) of the nucleon, which---in contrast to ChEFT---can also provide dynamical insights into the short-distance structure 
present in the generalized form factors.

This paper is organized as follows: In the next section we specify the operators with which we are going to obtain information on the three
generalized isoscalar and the three generalized isovector form factors of the nucleon. In section \ref{sec:formalism} we portray the effective
chiral Lagrangean required for the calculation, immediately followed by the sections containing our leading-one-loop 
results of the generalized
form factors in the isovector (section \ref{sec:isovector}) and in the isoscalar (section \ref{sec:isoscalar}) matrix elements. A summary of the 
main results concludes this paper, while a few technical details regarding the calculation of the amplitudes in covariant BChPT
are relegated to the appendices. 

\section{Extracting the First Moments of GPDs}
\label{sec:definitions}
\subsection{Generalized Form Factors of the Nucleon}

In Eqs.(\ref{momentsA},\ref{momentsB}) of the Introduction it was shown that the first moments of nucleon GPDs are connected to three generalized
form factors. In lattice QCD one can directly access the contribution of quark flavor $q$ to these generalized form factors of the nucleon 
by evaluating the matrix element \cite{reviews}
\begin{eqnarray}
 i\langle p'\vert\overline{q}\gamma_{\{ \mu}\overleftrightarrow{\textit{D}}_{\nu\}}\,q\vert p\rangle
&=& \overline{u}(p') \bigg[{\textit{A}_{2,0}^q (\Delta^2)}\gamma_{\{\mu}\overline{p}_{\nu\}}
- \frac{{\textit{B}_{2,0}^q (\Delta^2)} }{2M_{N}}\:\Delta^{\alpha}i\sigma_{\alpha\{\mu}\overline{p}_{\nu\}}
+\frac{ {\textit{C}_{2,0}^q (\Delta^2)} }{M_{N}}\:\Delta_{\{\mu}\Delta_{\nu\}}\bigg]\,  u(p). \label{basic}
\end{eqnarray}
The brackets $\{ \ldots \}$ denote the completely symmetrized and traceless combination of all indices in an operator. $u\,(\overline{u})$ is
a Dirac spinor of the incoming (outgoing) nucleon of mass $M_N$, for which the quark matrix-element is evaluated. In ChEFT we employ
the same philosophy and also extract information about the first moments of nucleon GPDs of Eq.(\ref{momentsA},\ref{momentsB}) via a 
calculation of the generalized form factors according to Eq.(\ref{basic}).

Studying a strongly interacting system with 2 light flavors in the non-perturbative regime of QCD, with the methods of ChEFT one can only 
{\em directly} access the singlet $(s)$ and triplet $(v)$ contributions of the quarks to the the three form factors:
\begin{eqnarray}
 i\langle p'\vert\overline{q}\gamma_{\{ \mu}\overleftrightarrow{\textit{D}}_{\nu\}}\,q\vert p\rangle_{u+d}
&=& \overline{u}(p') \bigg[{\textit{A}_{2,0}^s (\Delta^2)}\gamma_{\{\mu}\overline{p}_{\nu\}}
- \frac{{\textit{B}_{2,0}^s (\Delta^2)} }{2M_{N}}\:\Delta^{\alpha}i\sigma_{\alpha\{\mu}\overline{p}_{\nu\}}
+\frac{ {\textit{C}_{2,0}^s (\Delta^2)} }{M_{N}}\:\Delta_{\{\mu}\Delta_{\nu\}}\bigg]\, \frac{{\bf 1}}{2}\, u(p), \nonumber \\
& &\label{defs} \\
i\langle p'\vert\overline{q}\gamma_{\{ \mu}\overleftrightarrow{\textit{D}}_{\nu\}}\,q\vert p\rangle_{u-d}
&=& \overline{u}(p') \bigg[{\textit{A}_{2,0}^v (\Delta^2)}\gamma_{\{\mu}\overline{p}_{\nu\}}
- \frac{{\textit{B}_{2,0}^v (\Delta^2)} }{2M_{N}}\:\Delta^{\alpha}i\sigma_{\alpha\{\mu}\overline{p}_{\nu\}}
+\frac{ {\textit{C}_{2,0}^v (\Delta^2)} }{M_{N}}\:\Delta_{\{\mu}\Delta_{\nu\}}\bigg]\, \frac{\tau^a}{2}\, u(p). \nonumber \\
& & \label{defv}
\end{eqnarray}
Note that the shown 2 x 2 unit matrix ${\bf 1}$ and the Pauli-matrices $\tau^{a},\,a=1,2,3$ on the right hand sides of Eqs.(\ref{defs},\ref{defv}) 
operate in the space of a (proton,neutron) doublet field.

At present, not much is known\footnote{When adding on the contributions from gluons to the isoscalar quark matrix-elements of Eq.(\ref{defs}) 
one obtains $A_{2,0}^{s+g}(t=0)=1$ and $B_{2,0}^{s+g}(t=0)=0$ \cite{BJ}.} 
 yet experimentally about the momentum dependence of these 6 form factors. The main source of information at the moment
is provided by lattice QCD studies of these objects ({\it e.g.} see refs.\cite{QCDSF_dat,SESAM,LHPC}). Given that present-day lattice simulations work
with quark-masses much larger than those realized for $u$ and $d$ quarks in the standard model, one also needs to know the quark-mass
dependence of all 6 form factors, in order to extrapolate the lattice QCD results down to the real world of light $u$ and $d$ quarks. This
information is also encoded in the ChEFT results, typically expressed in form of a pion-mass dependence of the observables under study.
(The connection between the explicit breaking of chiral symmetry due to non-zero quark-masses and the resulting effective pion-mass is 
addressed in section \ref{meson}.) 

The need for a chiral extrapolation of lattice QCD results for the generalized form factors of the nucleon leads to a further complication in 
the analysis: One needs to be aware that it is {\it common practice in current lattice QCD analyses} that the mass
parameter $M_N$ in Eqs.(\ref{defs},\ref{defv}) does {\em not} correspond to the physical mass of a nucleon, instead, it represents a 
(larger) nucleon mass consistent with the values of the quark-masses employed in the simulation. 
Fortunately the quark-mass dependence of $M_N$ has been studied
in detail in ChEFT, both in non-relativistic \cite{NRmass} and in covariant frameworks \cite{Mass}. The next-to-leading one-loop chiral 
formulae of ref.\cite{Mass} provide a stable extrapolation function\footnote{Recent claims in the analysis of ref.\cite{Judith} reporting a possible 
breakdown of the chiral extrapolation formula for large pion masses at ${\cal O}(p^5)$ level in our opinion are due to an insufficient quark-mass
dependence of the  vertices entering the HBChPT formula at that order. This issue is under discussion \cite{GH}.} 
up to effective pion-masses $\sim 600$ MeV. In this work we utilize the ${\cal O}(p^4)$ 
BChPT result (see Appendix \ref{sec:amplitudes})
\begin{eqnarray}
M_N(m_{\pi}) & = & M_0-4c_1 m_{\pi}^2+\frac{3g_A^2m_{\pi}^3}{32\pi^2F_{\pi}^2\sqrt{4-\frac{m_{\pi}^2}{M_0^2}}}
	\left(-4+\frac{m_{\pi}^2}{M_0^2}+4c_1\frac{m_{\pi}^4}{M_0^3}\right)\arccos{\left(\frac{m_{\pi}}{2M_0}\right)}\nonumber \\
	& &+4e_1^r(\lambda)m_{\pi}^4-\frac{3m_{\pi}^4}{128\pi^2F_{\pi}^2}\Bigg[\left(\frac{6g_A^2}{M_0}-c_2\right)
       +4\left(\frac{g_A^2}{M_0}-8c_1+c_2+4c_3\right)\log{\left(\frac{m_{\pi}}{\lambda}\right)}\Bigg] \nonumber \\
& &-\frac{3c_1g_A^2m_{\pi}^6}{8\pi^2F_{\pi}^2M_0^2}\log{\left(\frac{m_{\pi}}{M_0}\right)}+{\cal O}(p^5). \label{p4mass}
\end{eqnarray}
in order to correct for the mass effects in Eqs.(\ref{defs},\ref{defv}). The coupling constants occurring in this formula are described in detail in 
ref.\cite{Mass}. Possible effects of higher orders can be estimated as ${\cal O}(p^5)\sim\delta_M\frac{m_\pi^5}{\Lambda_\chi^4}$, where 
$\delta_M$ could be varied within natural size estimates $-3,
\ldots,+3$ \cite{GH}.

We note that the trivial,  purely 
kinematical effect of $M_N=M_N(m_\pi)$ in Eqs.(\ref{defs},\ref{defv}) could induce quite a strong quark-mass dependence 
into the form factors $B_{2,0}^{s,v}(t),\,C_{2,0}^{s,v}(t)$ and might even be able to mask
any ``intrinsic'' quark-mass dependence in these form factors. We are reminded of the analysis of the Pauli form factors $F_2^{s,v}(t)$
in ref.\cite{QCDSFff}, where the absorption of the analogous effect into a ``normalized magneton'' even led to a different slope (!) 
for the isovector
anomalous magnetic moment $\kappa_v=F_2^v(t=0)$ when compared to the quark mass dependence of the ``not-normalized'' 
lattice data. We therefore
urge the readers that this effect should be taken into account in any quantitative (future) analysis of the quark-mass 
dependence of the generalized form factors
$B_{2,0}^{s,v}(t),\,C_{2,0}^{s,v}(t)$ as well. For completeness---based on the extensive studies of ref.\cite{Mass}---we suggest a set of 
couplings in table \ref{input} to be used in Eq.(\ref{p4mass}) from
which one can estimate the impact of this effect in the matrix elements
of Eqs.(\ref{defs}, \ref{defv}).

\begin{table}[tb]
\begin{center}
\begin{tabular}{|c|c|c|c|c|c|c|c|}
\hline
$g_A$ & $F_{\pi}$ [GeV] & $M_0$ [GeV] & $c_1$ [GeV$^{-1}$] & $c_2$ [GeV$^{-1}$] & $c_3$
[GeV$^{-1}$] & $e_1^r$(1GeV) [GeV$^{-3}$] & $\delta_M$ \\
1.2 & 0.0924 & 0.889 &-0.817 & 3.2 & -3.4 & 1.44 & 0\\
\hline
\end{tabular}
\caption{Input values used in this work for the numerical analysis of the chiral extrapolation functions.} \label{input}
\end{center}
\end{table}

Finally, we note that in the forward limit $t\rightarrow 0$ the generalized form factors $A_{2,0}^{s,v}(t=0)$ can be understood as moments
of the ordinary Parton Distribution Functions (PDFs) $q(x), \,\bar{q}(x)$ \cite{reviews}:
\begin{eqnarray}
\langle x\rangle_{u\pm d}&=&A_{2,0}^{s,v}(t=0)=\int_0^1 dx\,x\left(q(x)+\bar{q}(x)\right)_{u\pm d}. \label{defx}
\end{eqnarray}
Experimental results exist for $\langle x\rangle$ in proton- and ``neutron-'' targets, from which one can estimate the isoscalar and isovector
quark contributions at the physical point \cite{xproton} at a regularization scale $\mu$. In this work we choose $\mu=2$ GeV for our
comparisons with phenomenology\footnote{Note that this $\mu$-dependence is not part of the ChEFT framework, as it clearly involves
short-distance physics. However, all chiral couplings specified in section \ref{sec:formalism} carry an implicit $\mu$-dependence (which we do
not indicate), as soon as they are fitted to lattice QCD data or phenomenological values which do depend on this scale.}.  
In section \ref{t0} we attempt to connect the physical 
value for $\langle x\rangle_{u-d}$ with
recent (preliminary) lattice QCD results from the LHPC collaboration \cite{LHPC}, whereas in section \ref{sec:u+d} we analyse the quark-mass
dependence of $\langle x \rangle_{u+d}$ with (quenched) lattice QCD
results from the QCDSF collaboration \cite{QCDSF_dat}.

\subsection{Generalized Form Factors of the Pion}

The first moment of a generalized parton distribution function in a pion $\textit{H}^{q}_\pi(x,\xi,t)$ can be defined analogously to the case 
of a nucleon discussed above. One obtains \cite{reviews}
\begin{eqnarray}
\int_{-1}^{1}dx\,x\,\textit{H}^{q}_\pi(x,\xi,t)&=&A_{\pi}^{q}(t)+(-2\xi)^{2}\textit{C}_{\pi}^{q}(t). \label{pigpd}
\end{eqnarray}
The two functions $A_\pi^q(t),\,C_\pi^q(t)$ are the generalized form factors of the pion, generated by contributions of quark flavor $q$. 
In the forward limit one recovers the first moment of the ordinary parton distribution functions in a pion:
\begin{eqnarray}
\langle x\rangle_\pi&=&A_\pi^q(t=0)=\int_0^1 dx\,x\left(q(x)+\bar{q}(x)\right). \label{xpi}
\end{eqnarray}
In the analysis of the isoscalar GPD moments of a nucleon we encounter tensor fields directly interacting with the pion cloud of the nucleon.
One therefore needs to understand the relevant pion-tensor couplings in terms of the 2 generalized form factors $A_\pi(t),\,C_\pi(t)$.
We note that the two generalized form factors of the pion have been analysed at one-loop level already in ref.\cite{DL} for the total sum of 
quark and gluon contributions, whereas the quark contribution to the form factors as defined in Eq.(\ref{pigpd}) has been the focus of the 
more recent work \cite{Pion}.

\section{Formalism}
\label{sec:formalism} 
\subsection{Leading Order Nucleon Lagrangean}

The well-known leading order Lagrangean in BChPT is given as \cite{GSS}
\begin{equation}\label{LO}
\mathcal{L}_{\pi N}^{(1)}=\overline{\psi}_N\left\{i\gamma^\mu D_\mu - M_0 +\frac{g_A^0}{2}\gamma^\mu\gamma_5 u_\mu  \right\} \psi_N,
\end{equation}
with
\begin{eqnarray}
D_\mu \psi_N&=&\left\{ \partial_\mu-iv_{\mu}^{(0)} +\frac{1}{2}\left[u^\dagger,\partial_\mu u\right]-\frac{i}{2}u^\dagger\left(\tilde{v}_\mu+\tilde{a}_\mu\right) u 
-\frac{i}{2}u \left(\tilde{v}_\mu-\tilde{a}_\mu \right) u^\dagger \right\} \psi_N, \\
u_\mu&=&i u^\dagger \nabla_\mu U u^\dagger.
\end{eqnarray}
$U\equiv u^2$ corresponds to a non-linear realization of the (quasi-)
Goldstone boson fields, $\tilde{v}_\mu,\,\tilde{a}_\mu$ denote 
arbitrary isovector vector, axial-vector background fields, while
$v_{\mu}^{(0)}$ is the isosinglet vector background field. The covariant derivative $\nabla_\mu U$ is defined as
\begin{equation}
\nabla_\mu U = \partial_\mu U -i\left(\tilde{v}_\mu+\tilde{a}_\mu\right) U +i U \left(\tilde{v}_\mu-\tilde{a}_\mu\right).
\end{equation}
Finally, we note that the coupling $g_A^0$ denotes the axial-coupling of the nucleon (in the chiral limit), whereas $M_0$ corresponds to the 
nucleon mass (in the chiral limit).

We now want to extend this Lagrangean to the interaction between {\em external tensor fields} and a strongly interacting system at low 
energies. In this work we focus on {\em symmetric, traceless tensor fields} with positive parity in order to calculate the generalized form factors
of the nucleon. In particular, we utilize the (chiral) tensor structures
 \begin{eqnarray}
V_{\mu\nu}^\pm&=&\frac{1}{2} \left(g_{\mu\alpha}g_{\nu\beta}+g_{\mu\beta}g_{\nu\alpha}-\frac{2}{d}g_{\mu\nu}g_{\alpha\beta}\right)\times
                                   \left(u^\dagger V^{\alpha\beta}_R u  \pm u V^{\alpha\beta}_L u^\dagger \right),\nonumber \\
V_{\mu\nu}^0&=&\frac{1}{2}\left(g_{\mu\alpha}g_{\nu\beta}+g_{\mu\beta}g_{\nu\alpha}-\frac{2}{d}g_{\mu\nu}g_{\alpha\beta}\right)
\tilde{v}^{\alpha\beta} \; \frac{{\bf 1}}{2} \;.
\end{eqnarray}
 We note, that the chiral tensor fields $V_{\mu\nu}^\pm$ transform
as $h V_{\mu\nu}^\pm h^{\dagger}$ under chiral rotations, where $h$
corresponds to the standard compensator field of 2 flavour ChPT
\cite{GSS},
whereas $V_{\mu\nu}^0$ is a chiral singlet. Chiral transformation
properties for all remaining structures in our Lagrangeans are
standard and can be found in the literature.
The right- and left-handed fields $V_{\alpha\beta}^{(R,L)}$ are related to the symmetric (\textit{isovector}) tensor fields of definite parity 
$v_{\alpha\beta}^i,\, a_{\alpha\beta}^i,\, i=1,2,3$ via
\begin{eqnarray}
V_{\alpha\beta}^{R}&=&\left(v_{\alpha\beta}^i+a_{\alpha\beta}^i\right)\times\frac{\tau^i}{2}\, , \nonumber \\ 
V_{\alpha\beta}^{L}&=&\left(v_{\alpha\beta}^i-a_{\alpha\beta}^i\right)\times \frac{\tau^i}{2}\, ,
\end{eqnarray} 
while $\tilde{v}_{\alpha\beta}$ denotes the symmetric \textit{isoscalar} tensor field of positive parity.  

In order to study possible interactions with external tensor fields originating from the leading order Lagrangean Eq.(\ref{LO}), we re-write it into
the equivalent form
\begin{eqnarray}\label{im}
\mathcal{L}_{t\pi N}^{(0,1)}&=& \overline{\psi}_N\left\{\frac{1}{2} \left(i\gamma^\mu \widetilde{g_{\mu\nu}} \overrightarrow{D}^\nu-i
\overleftarrow{D}^\nu
\widetilde{g_{\mu\nu}} \gamma^\mu\right) - M_0 +\ldots\right\} \psi_N,
\end{eqnarray}
with
\begin{eqnarray}
\widetilde{g_{\mu\nu}}&=&g_{\mu\nu}+a_{2,0}^sV_{\mu\nu}^0+\frac{a_{2,0}^v}{2}V_{\mu\nu}^+.
\end{eqnarray}
The coupling $a_{2,0}^{s}\,(a_{2,0}^{v})$ has been defined such that it corresponds to the chiral limit value of 
$\langle x\rangle_{u-d}$ $(\langle x\rangle_{u+d})$ defined in Eq.(\ref{defx}). We note that the coupling $a_{2,0}^s$ is allowed to be different
from unity, as we only sum over the $u+d$ quark contributions in the isoscalar moments, but neglect the contributions from gluons. While this
separation between quark- and gluon-contributions does not occur in nature, it can be implemented in lattice QCD analyses at a fixed 
renormalization scale ({\it e.g} see refs.\cite{QCDSF_dat,SESAM,LHPC}). 

While the construction of the parity-even tensor interactions with a strongly interacting system started from the ${\cal O}(p)$ BChPT 
Lagrangean, inspection of the resulting Lagrangean Eq.(\ref{im}) reveals that the leading interactions actually start out at ${\cal O}(p^0)$, as
we do not assign a chiral power $p^n$ to any of the tensor fields.
Furthermore, symmetries allow the addition of the parity-odd tensor interaction $\sim V_{\mu\nu}^-$. We finally obtain
\begin{eqnarray}\label{p0}
\mathcal{L}_{t\pi N}^{(0)}&=& \frac{1}{2} \overline{\psi}_N \left\{i\gamma^\mu \left(a_{2,0}^sV_{\mu\nu}^0+\frac{a_{2,0}^v}{2}V_{\mu\nu}^+
+\frac{\Delta a_{2,0}^v}{2}V_{\mu\nu}^-\gamma_5\right)\overrightarrow{D}^\nu \right. \nonumber \\
& &\left. \quad\quad-i\overleftarrow{D}^\nu \gamma^\mu 
\left(a_{2,0}^sV_{\mu\nu}^0+\frac{a_{2,0}^v}{2}V_{\mu\nu}^++\frac{\Delta a_{2,0}^v}{2}V_{\mu\nu}^-\gamma_5\right) \right\} \psi_N,
\end{eqnarray}
with the coupling $\Delta a_{2,0}^v$ corresponding to the chiral limit value of the axial quantity $\langle\Delta x\rangle_{u-d}$ 
\cite{DHaxial}. 

The ${\cal O}(p^1)$ part of the leading order pion-nucleon Lagrangean in the presence of external symmetric, traceless tensor fields with 
positive parity then reads
\begin{eqnarray}
\mathcal{L}_{t\pi N}^{(1)}&=&\overline{\psi}_N\left\{i\gamma^\mu D_\mu - M_0 +\frac{g_A^0}{2}\gamma^\mu\gamma_5 u_\mu  
+\frac{b_{2,0}^v}{8 M_0}\bigg( i\sigma_{\alpha\mu}\:\left[\overrightarrow{D}^{\alpha},V^{\mu\nu}_+\right] \overrightarrow{D}_{\nu}+h.c.\bigg) \right.
\nonumber \\
& &\left.
\quad \quad+\frac{b_{2,0}^s}{4M_0}\bigg( i\sigma_{\alpha\mu}\:\left[\overrightarrow{\nabla}^{\alpha},V^{\mu\nu}_0\right] \overrightarrow{D}_{\nu}+h.c.\bigg)
+\ldots \right\} \psi_N, \label{p1}
\end{eqnarray}
where we have introduced $\nabla^{\alpha}=\partial^{\alpha}-iv_{(0)}^{\alpha}$.
The two new couplings $b_{2,0}^v,\,b_{2,0}^s$ can be interpreted as the chiral limit values of the form factors 
$B_{2,0}^v(0),\;B_{2,0}^s(0)$ in the limit $t\rightarrow 0$. No further structures enter our calculation at this order\footnote{We only show those 
terms where the tensor fields couple at tree level without simultaneous emission of
pions, photons, etc., as these are the relevant terms for our ${\cal O}(p^2)$ calculation of the form factors according to the power-counting
analysis of subsection \ref{counting}.}. Finally, we note that the coupling $b_{2,0}^s$ is only allowed to exist because we do not sum over the
gluon contributions in the isoscalar moments, otherwise this form factor is bound to vanish in the forward limit 
$B_{2,0}^{q+g}(t=0)\equiv 0$ \cite{anomgrav}.

\subsection{Consequences for the Meson Lagrangean}\label{meson}

The choice of assigning the chiral power $p^0$ to the symmetric tensor fields $V_{\mu\nu}^{L,R,0}$ also has the consequence that the
well-known leading-order chiral Lagrangean for 2 light flavours in the meson sector \cite{GL} is modified:
\begin{equation} \label{PiPi}
\mathcal{L}_{t\pi\pi}^{(2)}=\frac{F_0^2}{4}\; Tr\left[ \nabla_\mu U^\dagger \left(g^{\mu\nu}+4x_\pi^0 V^{\mu\nu}_0\right) 
\nabla_\nu U + \chi^\dagger U + \chi  U^\dagger \right].
\end{equation}
We note that the new coupling $x_\pi^0$ has been defined such that it corresponds to the chiral limit value of 
$\langle x\rangle_\pi$ of Eq.(\ref{xpi}). It is allowed to differ from unity because we only sum over the quark-distribution functions in the isoscalar
channel and neglect the contributions from gluons. 

The explicit breaking of chiral symmetry via the finite quark masses is encoded via
\begin{equation}
\chi=2\,B_0 \left(\tilde{s}+i\tilde{p}\right),
\end{equation}
if one switches off the external pseudoscalar background field $\tilde{p}$ and assigns the 2-flavour quark mass-matrix ${\it M_q}$ to the 
scalar background field $\tilde{s}$. To the order we are working here we obtain the resulting pion mass $m_\pi$ via
\begin{equation}
m_\pi^2=2\, B_0 \, \hat{m} + {\cal O}({\it M_q}^2),
\end{equation}
where $\hat{m}=(m_u+m_d)/2$ and $B_0$ corresponds to the value of the chiral condensate.
The other free parameter at this order $F_0$ can be identified with the value of the pion-decay constant (in the chiral limit).

\subsection{Power-counting with tensor fields in BChPT} \label{counting}

We start from the general power-counting formula of Baryon ChPT:
\begin{eqnarray}
D & = & 2L+1+\sum_{d}(d-2)N_d^M+\sum_d(d-1)N_d^{MB} \label{eq:pct}.
\end{eqnarray}
$D$ denotes the chiral dimension $p^D$ of a particular Feynman Diagram, $L$ counts the number of loops in the diagram, whereas
the variables $N_d^{M,\,MB}$ count the number of vertices of chiral dimension $d$ from the pion $(M)$ and pion-Nucleon $(MB)$ 
Lagrangeans.

To leading order we only have the tree level contributions from the $p^0$-Lagrangean of Eq.(\ref{p0}) with $L=0,\,N_2^M=0,\,N_0^{MB}=1$,
resulting in $D=0$.

At next-to-leading order $D=1$ we find additional tree level contributions from the $p^1$ Lagrangean of Eq.(\ref{p1}) with 
$L=0,\,N_2^M=0,\,N_1^{MB}=1$.

The first loop contributions enter at $D=2$ with $L=1,\,N_2^M=0,\,N_0^{MB}=1$ plus possible contributions from $N_1^{MB}$. The 
corresponding diagrams are shown in Fig.\ref{diags}. Diagram (e) in that Figure represents (loop) corrections from the nucleon Z-factor
(given in appendix \ref{sec:amplitudes}) which at this order renormalize only the tree-level tensor couplings of the $p^0$-Lagrangean. 
Note that there is an additional 
possibility of obtaining $D=2$ contributions via $L=0,\,N_2^M=0,\,N_2^{MB}=1$, corresponding to further tree level contributions discussed
in the next subsection. 

In this first paper we will stop with our analysis of the generalized form factors at the $D=2$, i.e. ${\cal O}(p^2)$ level,
corresponding to a (covariant) leading-one-loop calculation. The next-to-leading one-loop effects of $D=3$ will be postponed to a later
communication \cite{future}.

\begin{figure}[tb]
  \begin{center}
	\par\hspace{-2 cm}
   \includegraphics*[width=0.5\textwidth]{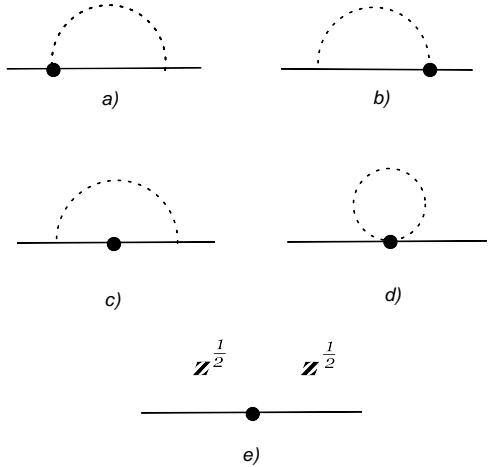}
    \caption{Loop diagrams contributing to the first moments of the
GPDs of a nucleon at leading-one-loop order in BChPT. The solid and
dashed lines represent nucleon and pion propagators respectively. The solid dot 
denotes a coupling to a tensor field from the ${\cal O}(p^0)$
Lagrangean of Eq.(\ref{p0}).  Aside from the trivial unity-contribution, wavefunction renormalization to the couplings of the order $p^1$
Lagrangean only start to contribute at next-to-leading one loop order.}
\label{diags}
  \end{center}
\end{figure}

\begin{figure}[tb]
  \begin{center}
   \includegraphics*[width=0.3\textwidth]{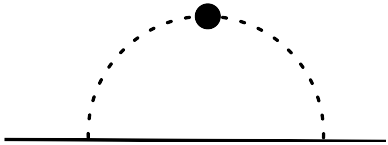}
    \caption{(Isoscalar) tensor field coupling to the pion cloud of the nucleon. This process only starts to contribute at next-to-leading one-loop
order in BChPT.}
\label{tensorpi}
  \end{center}
\end{figure}

The (perhaps) surprising finding of this powercounting analysis is the observation that the tensor coupling to the pion field controlled by the
coupling $x_\pi^0$ in Eq.(\ref{PiPi}) does {\em not} contribute at leading-one-loop order! It only starts to enter at $D=3$ via $L=1,\,N_2^M=1$
and $N_1^{MB}=1$ or $2$. (The corresponding diagram for $N_1^{MB}=2$ is shown in Fig.\ref{tensorpi}, while the not-shown diagram 
for $N_1^{MB}=1$ is expected to sum to zero due to isospin-symmetry.) We therefore note that the generalized form factors of the nucleon
behave quite different from the standard Dirac and Pauli form factors of the nucleon, where the pion-cloud interactions with the external source
a la Fig.\ref{tensorpi} are part of the leading-one-loop result and play a prominent role in the final result. We will discuss the impact of this
particular $D=3$ contribution further in section \ref{firstglance} when we try to estimate the possible size of higher order corrections to our ${\cal O}(p^2)$
analysis. 

We now move on to a discussion of the tensor interactions in the ${\cal O}(p^2)$ Lagrangean which are present at
$D=2$ according to our analysis.

\subsection{Next-to-leading Order Nucleon Lagrangean}
At next-to-leading order the covariant BChPT Lagrangean for 2-flavour QCD contains 7 independent terms in the presence of general scalar,
 pseudoscalar,
vector and axial-vector background fields, governed by the couplings\footnote{The couplings $c_6$ and $c_7$ are often denoted as 
$\kappa_v^0$ and $\kappa_s^0$ in the literature, as they constitute the leading terms in the isovector and isoscalar Pauli form factor 
of the nucleon \cite{GSS}.} $c_1, \ldots c_7$ \cite{BKM}. Extending this scenario to {\em symmetric} and {\em traceless} tensor background 
fields with positive parity, the symmetries allow the construction of {\em six} additional terms to describe the interaction at tree-level to 
this order\footnote{More 
couplings will appear when one wishes to extend this to more general scenarios.}:
\begin{equation}
\begin{split}
\mathcal{L}_{t\pi N}^{(2)}=
&\frac{c_8}{4M_0^2}\,\overline{\psi}_N\bigg\{ Tr(\chi_{+})  V_{\mu\nu}^+\gamma^{\mu}i\overrightarrow{D}^{\nu} +h.c.\bigg\} \psi_N \\
&+\frac{c_9}{2M_0^2}\,\overline{\psi}_N \bigg\{ Tr(\chi_{+})   \gamma^{\mu}i\overrightarrow{D}^{\nu} +h.c.\bigg\} \psi_N V_{\mu\nu}^0\\
&+\frac{c_{2,0}^v}{2M_0}\overline{\psi}_N\:\bigg\{\big[\overrightarrow{D}^{\mu},[\overrightarrow{D}^{\nu},V_{\mu\nu}^+]\big]\bigg\}\psi_N\\
&+\frac{c_{2,0}^s}{M_0}\overline{\psi}_N\:\bigg\{\big[\overrightarrow{\nabla}^{\mu},[\overrightarrow{\nabla}^{\nu},V_{\mu\nu}^0]\big]
\bigg\}\psi_N\\
&
+\frac{c_{12}}{4M_0^2}\overline{\psi}_N \bigg\{ \big[\overrightarrow{D}^{\alpha},[\overrightarrow{D}_{\alpha},
V_{\mu\nu}^+] \big] \gamma^{\mu}i\overrightarrow{D}^{\nu} +h.c. \bigg\} \psi_N\\
&
+\frac{c_{13}}{2M_0^2}\overline{\psi}_N \bigg\{ \gamma^{\mu}i\overrightarrow{D}^{\nu} +h.c. \bigg\} \psi_N 
\:\big[\overrightarrow{\nabla}^{\alpha},[\overrightarrow{\nabla}_{\alpha},V_{\mu\nu}^0] \big]\\
& + \ldots
\end{split}
\end{equation}
with $\chi_+=u^{\dagger}\chi u^{\dagger}+ u \chi^{\dagger}u$.
The physics behind these couplings $c_i,\,i=8, \ldots 13$ with respect to the generalized form factors of the nucleon
 is quite simple: $c_8\;(c_9)$ governs the leading quark-mass insertion in $\langle x\rangle_{u-d} \; (\langle x\rangle_{u+d})$, whereas
 the couplings $c_{10},\,c_{11}$ give the values of the generalized form factors $C_{2,0}^v(0),\,C_{2,0}^s(0)$ in the double limit 
$t\rightarrow 0,\,m_\pi\rightarrow 0$. We can therefore denote them as $c_{2,0}^v,\,c_{2,0}^s$. 
Finally, the couplings $c_{12}\,,c_{13}$ parametrize the contributions of {\em short-distance} physics to
 the slopes of the generalized form factors $A_{2,0}^v(t),\,A_{2,0}^s(t)$ (in the chiral limit). Note that the operator controlled by the coupling
$c_9$ is not allowed to exist when we add the gluon-contributions on the left hand side of Eq.(\ref{defs}) \cite{EMT}.\\
After laying down the necessary effective Lagrangeans for our calculation, we are now proceeding to the results of our calculation.

\section{Generalized isovector Form Factors in ${\cal O}(p^2)$ BChPT}
\label{sec:isovector} 
\subsection{Moments of the isovector GPDs at $t=0$} \label{t0}

\begin{figure}[tb]
  \begin{center}
   \includegraphics*[width=0.5\textwidth]{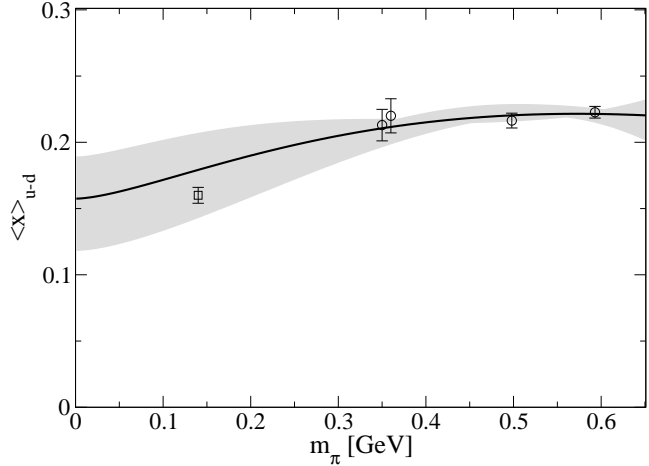}
    \caption{``Fit I'' of the ${\cal O}(p^2)$ result of Eq.(\ref{A20v0}) to the (preliminary) LHPC lattice data of ref.\cite{LHPC}. The 
corresponding parameters are given in Table \ref{table}. Note that the phenomenological value at physical pion mass was not included in 
the fit. The grey band shown indicates the estimate of possible ${\cal O}(p^3)$ corrections as discussed in the text.}
\label{x1}
  \end{center}
\end{figure}

In this subsection we present our results for the generalized
isovector form factors of the nucleon at $t=0$. For the PDF-moment $A_{2,0}^v(t=0)$
we obtain to ${\cal O}(p^2)$ in BChPT
\begin{eqnarray}
A_{2,0}^{v}(0) &\equiv& \langle x\rangle_{u-d} \nonumber \\
	&=&a_{2,0}^v+\frac{a_{2,0}^v\m^2}{(4\pi F_{\pi})^2}\Bigg\{-(3\g^2+1)\log\frac{\m^2}{\lambda^2}-2\g^2
	+\g^2\frac{\m^2}{\M^2}\bigg(1+3\log\frac{\m^2}{\M^2}\bigg) \nonumber \\
	&&  -\frac{1}{2}\g^2
	\frac{\m^4}{\M^4}\log\frac{\m^2}{\M^2}
	+\g^2 \frac{\m}{\sqrt{4\M^2-\m^2}}\bigg(14-8\frac{\m^2}{\M^2}+\frac{\m^4}{\M^4}\bigg)\arccos
	{\left(\frac{\m}{2\M}\right)}\Bigg\} \nonumber \\
	& &+\frac{\Delta a_{2,0}^v\g\m^2}{3(4\pi \F)^2}\Bigg\{2\frac{\m^2}{\M^2}\bigg(1+3\log\frac{\m^2}{\M^2}\bigg)-\frac{\m^4}{\M^4}\log\frac{\m^2}{\M^2}
	+\frac{2\m(4\M^2-\m^2)^{\frac{3}{2}}}{\M^4}\arccos\bigg(\frac{\m}{2\M}\bigg)\Bigg\} \nonumber \\
	& &+4\m^2 \frac{c_8^{(r)}(\lambda)}{M_0^2} + {\cal O}(p^3). \label{A20v0}
\end{eqnarray}
Many of the parameters in this expression are well known from analyses of chiral extrapolation functions. Numerical estimates for their chiral
limit values can be found in table \ref{input}. Furthermore, in a first fit to lattice data we constrain the coupling $\Delta a_{2,0}^v$ from the 
phenomenological value of $\langle \Delta x\rangle_{u-d}^{phen.}\approx 0.21$ via \cite{DHaxial}
\begin{equation}
\langle \Delta x\rangle_{u-d}=\Delta a_{2,0}^v + {\cal O}(m_\pi^2)
\end{equation} 
and perform a 2-parameter fit with the couplings 
$a_{2,0}^v,\,c_8^{(r)}(1\mbox{GeV})$ at the regularization scale $\lambda=1$ GeV. We fit to the (preliminary) LHPC data for this quantity as 
given in ref.\cite{LHPC}, including lattice data up to effective pion
masses of $m_{\pi} \approx$ 600 MeV. 
The resulting values for the fit parameters together with their statistical
errors are given in table \ref{table} 
and the resulting chiral extrapolation function is shown as the solid line in Fig.\ref{x1}. 
We note that the 
extrapolation curve tends towards smaller values for small quark-masses, but does not quite reach the phenomenological value at the 
physical point, which was {\em not} included in the fit. We have therefore tested what might happen if we estimate possible corrections to the 
solid curve arising from higher orders. From dimensional analysis we
know that the leading chiral contribution to $\langle x\rangle_{u-d}$ beyond our
calculation takes the form   ${\cal O}(p^3)\sim\delta_A
\,\frac{m_\pi^3}{\Lambda_\chi^2 M_0}+...$. Constraining 
$\delta_A$ between values\footnote{ The natural size of all   
couplings in the observables considered here is below 1, as all 
coupling estimates in this section refer to a moment of a parton 
distribution itself normalized to unity.
For this estimate we assume that the observable under consideration
has a well behaved chiral expansion, 
with the dominant contribution provided by the leading term in the chiral expansion.
This expectation is confirmed by the fit values of tables \ref{table} 
 and \ref{isoscalarfit} extracted for the investigated coupling constants.}
 $-1,\ldots,+1$  and 
repeating the fit with this uncertainty term included leads to the grey band 
indicated in Fig.\ref{x1}. 
Reassuringly, the phenomenological value for $\langle x\rangle_{u-d}$ lies well within that band of possible next-order corrections, giving us
{\em no indication} that something may be inconsistent with the large values for $\langle x\rangle_{u-d}$ typically found in lattice QCD
 simulations for large quark-masses. The resulting values for the couplings $a_{2,0}^v,\,c_8^{(r)}(1\mbox{GeV})$ of Fit I are also well within 
expectations. 

\begin{table}[tb]
\begin{center}
\begin{tabular}{|c|c|c|}
\hline
 & Fit I (4 points - 2 parameter) & Fit II (6+1 points - 3 parameter)
\\
\hline
$a_{2,0}^v$ & 0.157 $\pm$ 0.006 & 0.141 $\pm$ 0.0057 \\
$\Delta a_{2,0}^v$ & 0.210 (fixed)& 0.144 $\pm$ 0.034 \\
$c_8^r$(1GeV) & -0.283 $\pm$ 0.011& -0.213 $\pm$ 0.03 \\
\hline
\end{tabular}\\
\caption{Values of the couplings resulting from the two fits to the LHPC lattice data for $\langle x \rangle_{u-d}$. The errors shown are only
statistical and do neither include uncertainties from possible higher order corrections in ChEFT nor from systematic uncertainties connected with the 
lattice simulation.} \label{table}
\end{center}
\end{table}

We note that the mechanism of the downward-bending at small quark-masses in  $A_{2,0}^{v}(0)$ found in Eq.(\ref{A20v0}) is quite different
from what has been discussed so far in the literature within the non-relativistic HBChPT framework ({\it e.g.} see ref.\cite{detmold}). 
In order to demonstrate this we truncate Eq.(\ref{A20v0}) at $1/(16\pi^2F_\pi^2M_0)$ to obtain the exact ${\cal O}(p^2)$ HBChPT result of
refs.\cite{AS,CJ}:
\begin{eqnarray}
A_{2,0}^{v}(0)|_{HBChPT}^{p^2}&=&a_{2,0}^v\bigg\{1-\frac{\m^2}{(4\pi\F)^2}\bigg(2\g^2+(3\g^2+1)\log\frac{\m^2}{\lambda^2}\bigg)\bigg\}
+4\m^2 \frac{c_8^{(r)}(\lambda)}{M_0^2} \nonumber \\
& &\quad \quad+{\cal O}\left(\frac{1}{16\pi^2F_\pi^2M_0}\right). \label{xu-dHBChPT}
\end{eqnarray}
As already stated in the Introduction, the covariant BChPT scheme used in this work is able to reproduce exactly the corresponding 
non-relativistic HBChPT result at the same order by the appropriate truncation in $1/(16\pi^2F_\pi^2M_0)$. 

\begin{figure}[tb]
  \begin{center}
   \includegraphics*[width=0.5\textwidth]{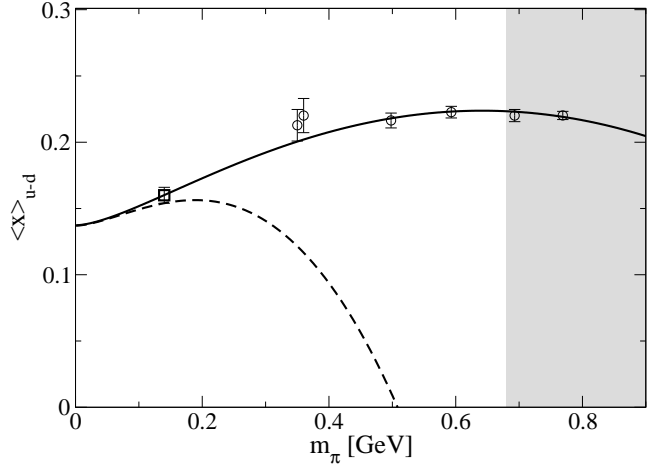}
    \caption{``Fit II'' of the ${\cal O}(p^2)$ BChPT result of Eq.(\ref{A20v0}) to the (preliminary) LHPC lattice data of ref.\cite{LHPC} {\em and} to the physical point 
(solid line). 
The corresponding fit-parameters are given in table \ref{table}. The dashed curve shown corresponds to ${\cal O}(p^2)$ result
in the HBChPT truncation (see Eq.(\ref{xu-dHBChPT})). The shaded area indicates the region where one does not expect that ChEFT can provide a
trustworthy chiral extrapolation function due to the large pion-masses involved, albeit the covariant ${\cal O}(p^2)$ result does not signal any 
breakdown in this region.}
\label{x2}
  \end{center}
\end{figure}

Fit I is certainly constricted by the assumption that we use the physical value of $\langle \Delta x\rangle_{u-d}^{phen.}\approx 0.21$ for the coupling 
$\Delta a_{2,0}^v$, which presumably takes a value in the chiral limit a bit smaller than the phenomenological value at the physical 
point \cite{DHaxial}. Furthermore, in order to also compare the ${\cal O}(p^2)$ HBChPT 
result of Eq.(\ref{xu-dHBChPT}) with the ${\cal O}(p^2)$ covariant BChPT result of Eq.(\ref{A20v0}) we perform a second fit: We fit the 
covariant expression for $\langle x\rangle_{u-d}$ of Eq.(\ref{A20v0}) again to the LHPC lattice data, 
this time, however, we constrain the coupling $\Delta a_{2,0}^v$ in such a way, that the resulting chiral extrapolation curve reproduces the 
phenomenological
value of $\langle x\rangle_{u-d}^{phen.}=0.160 \pm 0.006$ \cite{xproton} exactly for physical quark masses. The parameter values for this Fit II are again 
given in table \ref{table}, whereas
the resulting chiral extrapolation curve of the covariant ${\cal O}(p^2)$ expression of Eq.(\ref{A20v0}) is shown as the solid line in Fig.\ref{x2}. 
First, we would like to emphasize that the curve looks very reasonable, connecting the physical point with the lattice data of the LHPC
collaboration in a smooth fashion. Second, we note that the resulting values for the couplings $a_{2,0}^v,\, \Delta a_{2,0}^v$ underlying this curve 
are very 
reassuring, indicating that both $\langle x\rangle_{u-d}$ and $\langle \Delta x\rangle_{u-d}$ are slightly smaller in the chiral limit than at the 
physical point! Likewise, the unknown quark-mass insertion $c_8^{(r)}$ contributes in a strength just as expected from natural sizes 
estimates. For the comparison with HBChPT we now utilize the very same
values\footnote{ The covariant BChPT approach
applied in this work  guarantees that the chiral limit value for
an observables is mandated to be the same in both frameworks. For this
specific case this implies that there is one unique set of numerical
values to be determined for the couplings $a_{2,0}^v$ and $c_8^{(r)}$,
regardless of the ChEFT scheme employed.} for $a_{2,0}^v$ and $c_8^{(r)}$ of Fit II as given in table \ref{table}. The
resulting curve based on the ${\cal O}(p^2)$ HBChPT formula of Eq.(\ref{xu-dHBChPT})  is shown as the dashed curve in Fig.\ref{x2}. One
observes that this leading-one-loop HBChPT expression agrees with the covariant result between the chiral limit and the physical point, but
is not able to extrapolate on towards the lattice data\footnote{ We note that this behaviour of the 2-flavour HBChPT result is completely 
analogous to the corresponding
leading-one-loop HBChPT expressions for the axial coupling constant of the nucleon \cite{HPW} and also for the anomalous magnetic 
moments
of the nucleon \cite{HW}. It appears to be a pattern that such leading HBChPT extrapolation formulae only describe the quark-mass 
dependence 
between the chiral limit and the physical point. If one wants to push the chiral extrapolation to larger quark-masses, more 
$m_\pi$-dependencies seem to be required, which from the viewpoint of the HBChPT framework are ``higher order''.}.
 However, we note that all our {\em numerical} comparisons with HBChPT
results shown in the figures of 
section \ref{sec:isovector} and \ref{sec:isoscalar} are based on the assumption that the D=2 fit values
found in tables \ref{table} and \ref{isoscalarfit} are already reliable estimates of
 the true, correct values of these couplings in low
 energy QCD. Clearly, the shown HBChPT curves might
 have to be revised if future D=3 analyses \cite{future}
 lead to substantially different numerical values for
 these couplings. The true range of applicability of HBChPT
 versus covariant BChPT can only be determined, once the stability
 of the employed couplings is guaranteed. A study of higher order
 effects is therefore essential also in this respect.
Ideally we would like to reanalyse the results of Fit II by first fixing the
couplings from a fit of the HBChPT formula of Eq.(\ref{xu-dHBChPT}) and then
study the resulting $\mathcal{O}(p^2)$ BChPT curve for this
observable.  However, a fit to the present set of lattice data shown in
Fig.\ref{x1} leads to a curve that is not compatible with the 2 lightest
lattice points and lies significantly below the grey band shown in
Fig.\ref{x1}---even down to the chiral limit. We will study this issue
further in ref.\cite{future}.

 Keeping these caveats in mind, at this point we conclude that the smooth extrapolation behaviour of the covariant ${\cal O}(p^2)$ BChPT expression for $\langle x\rangle_{u-d}$ 
of Eq.(\ref{A20v0}) between the chiral limit and the region of present lattice QCD data is due to an {\em infinite tower} of 
$\left(\frac{m_\pi}{M_0}\right)^n$ terms.
According to our analysis  the chiral curvature resulting from the logarithm of Eq.(\ref{xu-dHBChPT}) governing the leading-non-analytic 
quark-mass behaviour of this moment is {\em not} responsible for the rising behaviour of the chiral extrapolation function, as had been hypothesized 
in ref.\cite{detmold}. It will be interesting to see how well our ${\cal O}(p^2)$ BChPT 
extrapolation formula for $\langle x\rangle_{u-d}$ will perform, once the new lattice QCD data from dynamical simulations at small quark 
masses for moments of GPDs become available \cite{latticenext}.  \\ 
Finally, we are discussing the ${\cal O}(p^2)$ BChPT results at $t=0$ for the remaining two generalized isovector form factors of the nucleon at twist-2
level. One obtains
\begin{eqnarray}
B_{2,0}^v(t=0)&=&b_{2,0}^v\frac{\Mn(\m)}{M_0}+
	\frac{a_{2,0}^v\, \g^2 \m^2}{(4\pi\F)^2}\,\bigg\{\bigg(3+\log\frac{\m^2}{\M^2}\bigg)-\frac{\m^2}{\M^2}
	\bigg(2+3\log\frac{\m^2}{\M^2}\bigg) \nonumber \\
	& &+\frac{\m^4}{\M^4}\log\frac{\m^2}{\M^2}-\frac{2\m}{\sqrt{4\M^2-\m^2}}\bigg(5-5\frac{\m^2}{\M^2}+\frac{\m^4}{\M^4}\bigg)
	\arccos\bigg(\frac{\m}{2\M}\bigg)\bigg\} +{\cal O}(p^3),
\label{B20vt0} 
\end{eqnarray}
\begin{eqnarray}
C_{2,0}^v(t=0)&=&c_{2,0}^v\frac{\Mn(\m)}{M_0}+\frac{a_{2,0}^v\g^2\m^2}{12(4\pi\F)^2}\,\bigg\{-1
	+2\frac{\m^2}{\M^2}\bigg(1+\log\frac{\m^2}{\M^2}\bigg)\nonumber \\&&-\frac{m_{\pi}^4}{M_0^4}\log{\frac{m_{\pi}^2}{M_0^2}}
	+\frac{2\m}{\sqrt{4\M^2-\m^2}}\bigg(2-4\frac{\m^2}{\M^2}+\frac{\m^4}{\M^4}\bigg)\arccos\bigg(\frac{\m}{2\M}\bigg)\bigg\} 
	+{\cal O}(p^3). \label{C20vt0}
\end{eqnarray}
As one can easily observe, one encounters plenty of non-analytic terms and even chiral logarithms in these covariant ${\cal O}(p^2)$ 
BChPT results. However, we note that in the HBChPT limit to the same chiral order ${\cal O}(p^2)$ one would only obtain the chiral limit
values $b_{2,0}^v,\,c_{2,0}^v$, nothing else \cite{Thesis}. E.g. the chiral logarithms 
calculated in ref.\cite{BJ} for $B_{2,0}^v(t=0),\,C_{2,0}^v(t=0)$ would only show up
in a full ${\cal O}(p^3)$, respectively ${\cal O}(p^4)$ BChPT calculation of the generalized form factors. From the viewpoint of power-counting
in BChPT they are to be considered part of the higher order corrections to the full ${\cal O}(p^2)$ results given in 
Eqs.(\ref{B20vt0},\ref{C20vt0}).
  
\begin{figure}[tb]
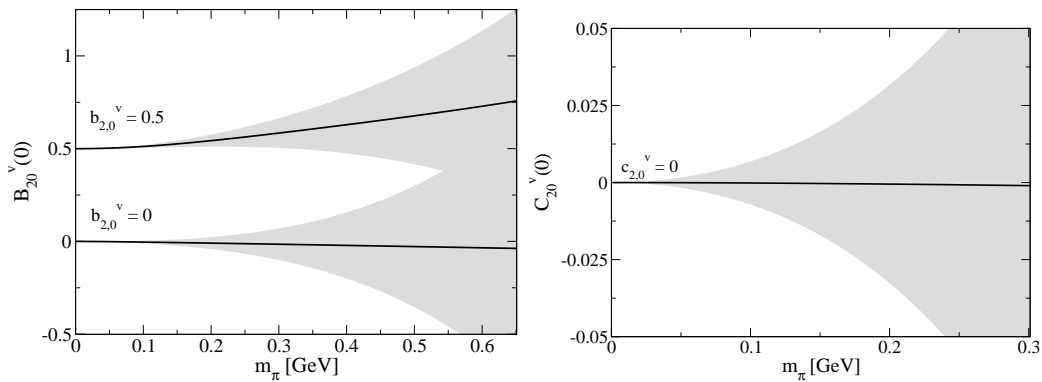

  \begin{center}
   \includegraphics*[width=0.4\textwidth]{B202.eps}
 \includegraphics*[width=0.4\textwidth]{C202.eps}
    \caption{Quark-mass dependence of the isovector moments $B_{2,0}^v(t=0)$ and $C_{2,0}^v(t=0)$. In $B_{2,0}^v(t=0)$ we have varied 
the (unknown) chiral limit value $b_{2,0}^v$ between $0$ and $+0.5$, as lattice analyses \cite{QCDSF_dat,SESAM,LHPC} suggest that this 
moment has a large positive value. For the chiral limit value
$c_{2,0}^v$ of $C_{2,0}^v(t=0)$ we have chosen the value 0, as
preliminary lattice QCD analyses
suggest that this moment is consistent with zero \cite{LHPC}. The grey bands shown indicate the size of possible higher order corrections to 
these ${\cal O}(p^2)$ results.} 
\label{PlotB20C20t0}
  \end{center}
\end{figure}

Unfortunately, at this point no information from experiment exists for these two structure quantities of the nucleon. From phenomenology one would expect that
$B_{2,0}^v(t=0)$ has a ``large'' positive value at $m_\pi=140$ MeV, as
it corresponds to the next-higher moment of the {\em isovector} Pauli form factor 
$F_2^v(t=0)\equiv\kappa_v=3.7$ n.m. (Compare Eq.(\ref{momentsB}) and Eq.(\ref{f2}) at $\xi=0$).
Lattice QCD analyses seem to support this expectation \cite{QCDSF_dat,SESAM,LHPC}. 
In contrast, the value of $C_{2,0}^v(t=0)$ cannot be estimated from information known about nucleon
form factors. State-of-the-art lattice QCD analyses ({\it e.g.} see ref.\cite{LHPC}) suggest that it is consistent with zero. In 
Fig.\ref{PlotB20C20t0} we have indicated how the corresponding extrapolation curves based upon this information might look like. We are looking forward to
a combined analysis of Eqs(\ref{B20vt0},\ref{C20vt0}) and the new lattice QCD data from dynamical simulations at small quark-masses \cite{latticenext} 
extrapolated to $t=0$, which will shed new light onto this new domain of nucleon structure physics!
As a caveat we note that in both form factors we would see a dominant influence of the quark-mass dependence stemming from the kinematical 
factor $M_N(m_\pi)$ in Eq.(\ref{defv}), {\em if} their corresponding chiral limit values $b_{2,0}^v,\,c_{2,0}^v\neq 0$. 
A further observation is that the uncertainties connected with possible higher order corrections from 
${\cal O}(p^3)$ could already become substantial for pion masses
around 300 MeV. For both quantities they can be estimated
via\footnote{Note, that the momentum $\Delta_{\mu}$ in
Eqs.(\ref{defs},\ref{defv})
 counts itself as an object of order $p$ leading to the
observation that a $D=3$ evaluation of the currents Eqs.(\ref{defs},\ref{defv})
leads to a leading quark-mass dependence of $m_{\pi}^{D-1}$ in
$B_{2,0}^{(s,v)}(0)$.}
${\cal O}(p^3)\sim \delta_{B,C}\frac{m_\pi^2 M_N(m_\pi)}{\Lambda_\chi^2 M_0}$ where $-1<\delta_{B,C}<1$. In order to ultimately test the stability of the results in 
Eqs.(\ref{B20vt0},\ref{C20vt0}) it will be very useful to extend this analysis to next-to-leading one-loop order \cite{future}.

\subsection{The slopes of the generalized isovector Form Factors}

In order to discuss the generalized isovector form factors $A_{2,0}^v(t),\,B_{2,0}^v(t),\,C_{2,0}^v(t)$ at non-zero values of $t$, we first want to 
analyze their slopes $\rho_X$, defined via
\begin{eqnarray}
X_{2,0}^v(t)&=&X_{2,0}^v(0)+\rho_X^v\,t \; + {\cal O}(t^2);\quad\quad X=A,B,C.
\end{eqnarray}
To ${\cal O}(p^2)$ in BChPT we find
\begin{eqnarray}
\rho_A^v &=&
	\frac{c_{12}}{\M^2}-\frac{a_{2,0}^v
	\g^2}{6(4\pi\F)^2}\,\frac{\m^2}{(4\M^2-\m^2)}\,\bigg\{26+8\log\frac{\m^2}{\M^2}
	-\frac{\m^2}{\M^2}\bigg(30+32\log\frac{\m^2}{\M^2}\bigg)\nonumber\\&&+\frac{\m^4}{\M^4}\bigg(6+\frac{39}{2}\log\frac{\m^2}{\M^2}\bigg)
	-3\frac{\m^6}{\M^6}\log\frac{\m^2}{\M^2}\nonumber
	\\&&
	-\frac{\m}{\sqrt{4\M^2-\m^2}}\bigg(90
	-130\frac{\m^2}{\M^2}+51\frac{\m^4}{\M^4}-6\frac{\m^6}{\M^6}\bigg)\arccos\bigg(\frac{\m}{2\M}\bigg)\bigg\}+\frac{m_{\pi}}{\Lambda_{\chi}^2M_0}\delta_A^t,\label{sla}
\end{eqnarray}
\begin{eqnarray}
\rho_B^v&=&
	\frac{a_{2,0}^v \g^2}{18(4\pi\F)^2} \,\frac{1}{(4\M^2-\m^2)}\bigg\{4\M^2+\m^2\bigg(83+24\log\frac{\m^2}{\M^2}\bigg)
	-114\frac{\m^4}{\M^2}\bigg(1+\log\frac{\m^2}{\M^2}\bigg)\nonumber \\
	& &+\frac{\m^6}{\M^4}\bigg(24+75\log\frac{\m^2}{\M^2}\bigg)-12\frac{\m^8}{\M^6}\log\frac{\m^2}{\M^2}
	-\frac{6\m^3}{\sqrt{4\M^2-\m^2}}\bigg(50-80\frac{\m^2}{\M^2}+33\frac{\m^4}{\M^4}-4\frac{\m^6}{\M^6}\bigg) \nonumber \\
	& &\times \arccos\bigg(\frac{\m}{2\M}\bigg)\bigg\} +\delta_B^t\,\frac{\Mn (m_\pi)}{\Lambda_\chi^2 M_0}, \\
\rho_C^v&=&
	\frac{a_{2,0}^v \g^2 }{180 (4\pi\F)^2}\frac{1}{(4\M^2-\m^2)}\bigg\{4\M^2-13 \m^2
	+12 \frac{\m^4}{\M^2}\bigg(4+3 \log\frac{\m^2}{\M^2}\bigg)\nonumber \\
	& &-3\frac{\m^6}{\M^4}\bigg(4+11\log\frac{\m^2}{\M^2}\bigg)+6\frac{\m^8}{\M^6}\log\frac{\m^2}{\M^2}
	+\frac{6\m^3}{\sqrt{4\M^2-\m^2}}\bigg(10-30\frac{\m^2}{\M^2}+15\frac{\m^4}{\M^4}-2\frac{\m^6}{\M^6}\bigg)\nonumber \\
	& &\times \arccos\bigg(\frac{\m}{2\M}\bigg)\bigg\} +\delta_C^t\,\frac{\Mn (m_\pi)}{\Lambda_\chi^2 M_0}. \label{slc} 
\end{eqnarray}
The parameters $\delta_A^t,\,\delta_B^t,\,\delta_C^t$ are not part of
the covariant $\mathcal{O}(p^2)$ result. They are only given to indicate the size of possible higher order corrections from ${\cal O}(p^3)$. 
A numerical analysis of the formulae given above suggests that the size
of pion-cloud contributions to the slopes of the {\em generalized isovector} form factors is very small! The physics governing the size
of these objects seems to be hidden in the counterterm contributions $c_{12},\,\delta_B^t,\,\delta_C^t$ which dominate numerically 
when assuming
natural size estimates $-1<c_{12},\,\delta_B^t,\,\delta_C^t<+1$. We note that this situation reminds us
 of the {\em isoscalar Dirac and Pauli} form factors of the nucleon $F_1^s(t),\,F_2^s(t)$, 
where the t-dependence in SU(2) ChEFT calculations is also dominated by counterterms ({\it e.g.} see the discussion in ref.\cite{BFHM}).  

Finally, truncating the covariant results of Eqs.(\ref{sla}-\ref{slc}) 
at $1/(16\pi^2F_\pi^2M_0)$ we obtain the (trivial) HBChPT expressions to ${\cal O}(p^2)$:
\begin{eqnarray}
\rho_A^v&=&\frac{c_{12}}{M_0^2} + {\cal O}(p^3), \\
\rho_B^v&=&0+ {\cal O}(p^3), \\
\rho_C^v&=&0+ {\cal O}(p^3). 
\end{eqnarray}
The non-zero slope results found in the HBChPT calculations of refs.\cite{ACK,DM} are therefore of higher order from the point
of view of our powercounting. Most of them can already be added systematically to our covariant ${\cal O}(p^2)$ results of Eqs.(\ref{sla}-\ref{slc}) at 
${\cal O}(p^3)$ \cite{future}.

\subsection{Generalized isovector Form Factors of the Nucleon }

In this subsection we present the full t-dependence of the (generalized) isovector form factors of a nucleon to ${\cal O}(p^2)$ in BChPT. 
We note that for all three form factors at this order only diagram c) of Fig.\ref{diags} gives a non-zero contribution at finite values of $t$---the 
resulting expressions at this order are therefore quite simple: 
\begin{eqnarray}
A_{2,0}^v(t) & = & A_{2,0}^v(0)+\frac{a_{2,0}^v g_A^2}{192 \pi^2 F_{\pi}^2} F_{2,0}^v(t)+\frac{c_{12}}{M_0^2}\,t+\mathcal{O}(p^3), \label{A20v}
\end{eqnarray}
with
\begin{eqnarray}
F_{2,0}^v(t) & = & t-\frac{2m_{\pi}^3
	\left(50M_0^4-43m_{\pi}^2M_0^2+8m_{\pi}^4\right)}{M_0^4\sqrt{4M_0^2-m_{\pi}^2}}
	\arccos{\left(\frac{m_{\pi}}{2M_0}\right)} \nonumber \\
&&+\int_{-\frac{1}{2}}^{\frac{1}{2}}\,du\Bigg\{\frac{2m_{\pi}^3
	M_0^2}{\tilde{M}^8\sqrt{4\tilde{M}^2-m_{\pi}^2}}
	\left[-10\tilde{M}^6+\left(17m_{\pi}^2+60M_0^2\right)\tilde{M}^4-4m_{\pi}^2\left(m_{\pi}^2+15M_0^2\right)\tilde{M}^2\right. \nonumber \\
&&\left.+12m_{\pi}^4M_0^2\right]
	\arccos{\left(\frac{m_{\pi}}{2\tilde{M}}\right)}+\frac{2\left(M_0^2-\tilde{M}^2\right)}{M_0^2\tilde{M}^6}
	\left[12m_{\pi}^4M_0^4+2m_{\pi}^2\left(4m_{\pi}^2-9M_0^2\right)M_0^2\tilde{M}^2\right.\nonumber \\
&&\left.+\left(m_{\pi}^2-2M_0^2\right)\left(8m_{\pi}^2+M_0^2\right)\tilde{M}^4
	\right]+\frac{2M_0^2}{\tilde{M}^8}\left[2\tilde{M}^8+3m_{\pi}^2\left(3m_{\pi}^2+4M_0^2\right)\tilde{M}^4\right. \nonumber \\
&&\left.-4m_{\pi}^4
	\left(m_{\pi}^2+9M_0^2\right)\tilde{M}^2+12m_{\pi}^6M_0^2\right]\log{\left(\frac{\tilde{M}}{M_0}\right)} +\frac{2m_{\pi}^2}{M_0^4
            \tilde{M}^8}\left[m_{\pi}^4\left(-12M_0^8+4\tilde{M}^2M_0^6\right.\right. \nonumber \\
&&\left. \left.+8\tilde{M}^8\right)+9m_{\pi}^2\tilde{M}^2
	\left(4M_0^8-\tilde{M}^2M_0^6-3\tilde{M}^6M_0^2\right)+12M_0^4\tilde{M}^4\left(\tilde{M}^4-M_0^4\right)\right]\log{\left(
            \frac{m_{\pi}}{M_0}\right)}\Bigg\}, \nonumber \\
&&\label{davt}
\end{eqnarray}
and $\tilde{M}^2=\M^2+\left(u^2-\frac{1}{4}\right)t$. Note that $F_{2,0}^v(t=0)\equiv 0$, whereas $A_{2,0}^v(0)$ has been discussed in 
section \ref{t0}. A conservative estimate for the size of possible higher order corrections indicated in 
Eq.(\ref{A20v})
can be obtained via $\mathcal{O}(p^3)\sim \delta_A \,\frac{m_\pi^3}{\Lambda_\chi^2 M_0}+\delta_A^t\,\frac{m_\pi}{\Lambda_\chi^2 M_0}\,t$, as discussed 
in the two
previous subsections. For the remaining two (generalized) isovector form factors we obtain
\begin{eqnarray}
B_{2,0}^v(t) & = & b_{2,0}^v\,\frac{\Mn(m_{\pi})}{\M}+\frac{a_{2,0}^v
	g_A^2M_0^2}{48\pi^2F_{\pi}^2}\int_{-\frac{1}{2}}^{\frac{1}{2}}\!
	\frac{du}{\tilde{M}^8}\Bigg\{\left(M_0^2-\tilde{M}^2\right)\tilde{M}^6+9m_{\pi}^2M_0^2\tilde{M}^4\nonumber \\ && -6m_{\pi}^4M_0^2\tilde{M}^2+6m_{\pi}^2M_0^2
	\left(m_{\pi}^4-3m_{\pi}^2\tilde{M}^2+\tilde{M}^4\right)\log{\frac{m_{\pi}}{\tilde{M}}}\nonumber \\ &&-\frac{6m_{\pi}^3M_0^2}{\sqrt{4\tilde{M}^2-m_{\pi}^2}}\bigg[
	m_{\pi}^4-5m_{\pi}^2\tilde{M}^2+5\tilde{M}^4\bigg]\arccos{\left(\frac{m_{\pi}}{2\tilde{M}}\right)}\Bigg\}\nonumber \\ &&+\delta_B
	\,\frac{m_\pi^2M_N(m_\pi)}{\Lambda_\chi^2 M_0}+\delta_B^t\,\frac{M_N(m_\pi)}{\Lambda_\chi^2 M_0}\,t,
\end{eqnarray}
\begin{eqnarray}
C_{2,0}^v(t) & = & c_{2,0}^v\frac{M_N(m_\pi)}{M_0}+
	\frac{a_{2,0}^vg_A^2M_0^2}{48\pi^2F_{\pi}^2}\int_{-\frac{1}{2}}^{\frac{1}{2}}\frac{du\,u^2}{\tilde{M}^8}\Bigg\{2\left(M_0^2-\tilde{M}^2\right)\tilde{M}^6
	-3m_{\pi}^2M_0^2\tilde{M}^4\nonumber \\&&+6m_{\pi}^4M_0^2\tilde{M}^2-6m_{\pi}^4M_0^2\left(m_{\pi}^2-2\tilde{M}^2\right)\log{\frac{m_{\pi}}{\tilde{M}}}
	\nonumber
	\\&&+\frac{6m_{\pi}^3M_0^2}{\sqrt{4\tilde{M}^2-m_{\pi}^2}}\bigg[m_{\pi}^4-4m_{\pi}^2\tilde{M}^2+2\tilde{M}^4\bigg]
	\arccos{\left(\frac{m_{\pi}}{2\tilde{M}}\right)}\Bigg\}
	\nonumber \\&&+\delta_C \,\frac{m_\pi^2M_N(m_\pi)}{\Lambda_\chi^2M_0}+\delta_C^t\,\frac{M_N(m_\pi)}{\Lambda_\chi^2 M_0}\,t.
\end{eqnarray}
The parameters $\delta_B,\,  \delta_C,\, \delta_B^t,\,  \delta_C^t $ have been inserted to study the possible corrections of higher orders 
to our ${\cal O}(p^2)$ results. Varying
these parameters between -1 and +1 we 
conclude that
a full ${\cal O}(p^3)$ calculation is required, before we want to make any strong claims regarding the t-dependence of $B_{2,0}^v(t)$ and  
$C_{2,0}^v(t)$ {\em beyond} the linear t-dependence discussed in the previous subsection. 
On the other hand, the predicted t-dependence of the form factor $A_{2,0}^v(t)$ of Eq.(\ref{A20v}) appears to be more reliable
at this order, as possible higher order contributions only affect terms beyond the leading linear dependence in t. 

\section{Generalized isoscalar Form Factors in ${\cal O}(p^2)$ BChPT}
\label{sec:isoscalar}
\subsection{Moments of the isoscalar GPDs at $t=0$}\label{sec:u+d}

To ${\cal O}(p^2)$ in 2-flavor covariant BChPT the only non-zero loop contributions to the isoscalar moment $A_{2,0}^s(t=0)$ (see Eq.(\ref{defx})) 
arise from diagrams c) and e) in Fig.\ref{diags}. One obtains
\begin{eqnarray}
A_{2,0}^s(0)&=&\langle x\rangle_{u+d} \nonumber \\
        &=& a_{2,0}^s+4m_{\pi}^2\frac{c_9}{M_0^2}
	-\frac{3a_{2,0}^sg_A^2 m_{\pi}^2}{16\pi^2 F_{\pi}^2}\Bigg[\frac{\m^2}{M_0^2}+\frac{m_{\pi}^2}{M_0^2}
       	\left(2-\frac{m_{\pi}^2}{M_0^2}\right)\log{\left(\frac{m_{\pi}}{M_0}\right)} \nonumber \\
        &  & + \frac{m_{\pi}}{\sqrt{4M_0^2-m_{\pi}^2}}\left(2-4\frac{m_{\pi}^2}{M_0^2}+\frac{m_{\pi}^4}{M_0^4}\right)
        \arccos{\left(\frac{m_{\pi}}{2M_0}\right)}\Bigg]+{\cal O}(p^3). \label{xs}
\end{eqnarray}
Eq.(\ref{xs}) should provide a similarly successful chiral extrapolation function for $\langle x\rangle_{u+d}$ as the 
covariant ${\cal O}(p^2)$ BChPT result of Eq.(\ref{A20v0}) did
for the LHPC lattice data for $\langle x\rangle_{u-d}$ in section \ref{t0}. The uncertainty arising from higher orders can be estimated to
scale as ${\cal O}(p^3) \sim \delta_A^0\frac{m_\pi^3}{\Lambda_\chi^2 M_0}$, where $\delta_A^0$ should be a number between $-1,\ldots,+1$, 
according to natural size
estimates. We note that the coupling $\Delta a_{2,0}^v$, which played an essential role in the chiral extrapolation function of 
$\langle x\rangle_{u-d}$,
is \textit{not} present in the quark-mass dependence of the isoscalar moment $\langle x\rangle_{u+d}$. The resulting chiral extrapolation function is
therefore presumably quite different from the one in the isovector channel. The absence of a chiral logarithm $\sim m_\pi^2\log m_\pi$ in 
$\langle x\rangle_{u+d}$ (compare Eq.(\ref{xu-dHBChPT}) and   Eq.(\ref{xs})) presumably will only lead to a difference in the chiral 
extrapolation functions between the isovector and the isoscalar moment for values $m_\pi<140$ MeV.  

Note that from Eq.(\ref{xs}) in the limit $1/(16\pi^2F_\pi^2M_0)\rightarrow 0$ we reproduce the leading HBChPT result for 
$\langle x\rangle_{u+d}$ of ref.\cite{AS}, which found a complete cancellation of the 
non-analytic quark-mass dependent terms in this channel (in the static limit):
\begin{eqnarray}
 A_{2,0}^s(0)|_{HBChPT}^{p^2}&=& a_{2,0}^s+4m_{\pi}^2\frac{c_9}{M_0^2}+{\cal O}(1/(16\pi^2F_\pi^2M_0)).  \label{xsHB}
\end{eqnarray}
The coupling $c_9$ is therefore scale independent (in 
dimensional regularization)
and constitutes the leading correction to the chiral limit value $a_{2,0}^s$ of $\langle x\rangle_{u+d}$. 

At $t=0$ we also find non-trivial results for the two other generalized isoscalar form factors of the nucleon to this order in the calculation
\begin{eqnarray}
B_{2,0}^s(0)&=&b_{2,0}^s\,\frac{\Mn(m_\pi)}{M_0}-
	\frac{3 \,a_{2,0}^s\, \g^2 \m^2}{(4\pi\F)^2}\,\bigg\{\bigg(3+\log\frac{\m^2}{\M^2}\bigg)-\frac{\m^2}{\M^2}\bigg(2+3\log\frac{\m^2}{\M^2}\bigg) 
         \nonumber\\
	&&+\frac{\m^4}{\M^4}\log\frac{\m^2}{\M^2}-\frac{2\m}{\sqrt{4\M^2-\m^2}}\bigg(5-5\frac{\m^2}{\M^2}+\frac{\m^4}{\M^4}\bigg)
	\arccos\bigg(\frac{\m}{2\M}\bigg)\bigg\} \nonumber \\
	& &+\delta_B^0\frac{m_\pi^2M_N(m_\pi)}{\Lambda_\chi^2 M_0}, \label{B20seq}\\
C_{2,0}^s(0)&=&c_{2,0}^s\,\frac{\Mn(m_\pi)}{M_0}-\frac{\,a_{2,0}^s\g^2\m^2}{4(4\pi\F)^2}\,\bigg\{-1
	+2\frac{\m^2}{\M^2}\bigg(1+\log\frac{\m^2}{\M^2}
	\bigg)-\frac{\m^4}{\M^4}\log\frac{\m^2}{\M^2} \nonumber \\
	& &+\frac{2\m}{\sqrt{4\M^2-\m^2}}\bigg(2-4\frac{\m^2}{\M^2}+\frac{\m^4}{\M^4}\bigg)\arccos\bigg(\frac{\m}{2\M}\bigg)\bigg\}
	+\delta_C^0\frac{m_\pi M_N(m_\pi)}{\Lambda_\chi^2}.
	\label{c20seq}
\end{eqnarray}
Note that additional chiral logarithms appear in the HBChPT
calculations of refs.\cite{ACK,DM}, {\it e.g.} a term $~b_{20}^s m_{\pi}^2 \log{m_{\pi}}$ entering the analogue of 
Eq.(\ref{B20seq}). According to the chiral power-counting formula
Eq.(\ref{eq:pct}) such a term only starts to appear at
$\sim\mathcal{O}(p^3)$, and is therefore not present in our $\mathcal{O}(p^2)$ calculation.
We have added the parameters $\delta_B^0,\,\delta_C^0$  ``by
hand'' to our covariant $\mathcal{O}(p^2)$ results Eqs.(\ref{B20seq})
and (\ref{c20seq}) in order to
indicate possible effects of higher order ({\it i.e.} ${\cal O}(p^3)$)
corrections. Presumably they take
on values $-1,\ldots,+1$. Note that our ${\cal O}(p^2)$ BChPT prediction for $C_{2,0}^s(0)$ is strongly affected by possible corrections from higher orders. This is due
to the fact that one only receives a non-zero result for this form factor starting at ${\cal O}(p^2)$---the order we are working in. Eq.(\ref{c20seq}) should therefore
only be considered to provide a rough estimate for the quark mass dependence of this form factor at $t=0$. For a true quantitative analysis of its chiral extrapolation
behaviour the complete\footnote{The most prominent correction from ${\cal O}(p^3)$ arises from the triangle diagram of Fig.\ref{tensorpi} and
is given in appendix \ref{sec:isosp3} via $\Delta C_{h.o.}^s(t=0,m_\pi)$. We note, however, that this is not the only correction at the next order.}
 ${\cal O}(p^3)$ corrections should first be added \cite{future}.

\begin{figure}[tb]
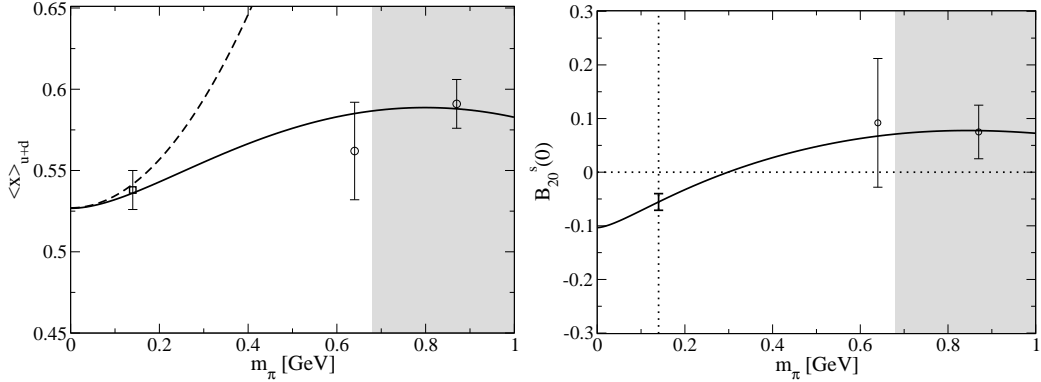

  \begin{center}
   \includegraphics*[width=0.4\textwidth]{A20s.eps}   \includegraphics*[width=0.4\textwidth]{B20s2.eps}
    \caption{Solid lines: The quark mass dependence of the  $\mathcal{O}(p^2)$ BChPT result for $\langle
x\rangle_{u+d}$ of Eq.(\ref{xs}) and $B_{2,0}^s(0)$ of
Eq.(\ref{B20seq}). The three unknown parameters resulting from a
{\em combined} fit to the shown QCDSF data of ref.\cite{QCDSF_dat} and to the phenomenological value for 
$\langle x\rangle_{u+d}$ \cite{xproton} are given in table \ref{isoscalarfit}. At the physical point we obtain 
$B_{2,0}^s=-0.056\pm 0.016$, with the shown error only being statistical. Due to the
poor data situation with its unknown systematic errors, however,  these results should only be considered as a rough estimate of the true quark-mass dependence. 
 The dashed line corresponds to the respective HBChPT
result at this order.} 
\label{Plotab}
  \end{center}
\end{figure}

Unfortunately, the (published) lattice QCD data base for $\langle x\rangle_{u+d}$ and $B_{2,0}^s(t\rightarrow 0)$ is quite sparse for 
small values of the pion mass. 
In order to obtain a {\em rough estimate} of the chiral extrapolation functions resulting from Eqs.(\ref{xs},\ref{B20seq}) we have to resort to 
(quenched)\footnote{The quark-masses employed in ref.\cite{QCDSF_dat} are so large, that one does not expect to find
differences between quenched and dynamical simulations. See {\it e.g.} the discussion ref.\cite{HW}. Note that we are utilizing the lattice data of 
ref.\cite{QCDSF_dat} with
the scale set by $r_0$, as we consider the alternative way of scale-setting (via a {\it linear} extrapolation to the physical mass of the nucleon) also discussed in 
ref.\cite{QCDSF_dat} to be obsolete in the light of the detailed
chiral extrapolation studies of ref.\cite{Mass}.
 We further note that in the simulation results 
of ref.\cite{QCDSF_dat} all contributions from disconnected-diagrams have been neglected, which gives rise to an additional unknown systematic uncertainty in the lattice 
data.}
 data of the QCDSF collaboration in 
ref.\cite{QCDSF_dat}. Performing a combined fit of Eqs.(\ref{xs},\ref{B20seq})
to the lattice data shown in Fig.\ref{Plotab} and including the phenomenological value of $\langle x\rangle_{u+d}\sim 0.54$ \cite{xproton} we obtain 
the two solid curves shown in 
Fig.\ref{Plotab}. The resulting parameters of the fit are given in table \ref{isoscalarfit}. Interestingly, despite the large quark-masses and the 
huge error bars in the
data of ref.\cite{QCDSF_dat} we obtain reasonable chiral extrapolation curves with natural size couplings. The analysis of the QCDSF data
in combination with the physical value for $\langle x\rangle_{u+d}$ suggests that the chiral limit value of this isoscalar PDF-moment is smaller
than the value at the physical point, leading to a monotonically rising chiral extrapolation function as shown in the left panel of Fig.\ref{Plotab}. 
It will be very interesting to observe what values an upcoming analysis of the new LHPC data
will find \cite{latticenext} for this moment, as a glance at the {\em preliminary} results shown in ref.\cite{LHPC} seems to indicate that for effective pion masses
around 300 MeV LHPC finds values for 
$\langle x\rangle_{u+d}$ which are {\em lower} than the physical point. If this observation can be confirmed in the final analysis of the data 
\cite{latticenext} one would have to conclude that the chiral extrapolation behaviour of $\langle x\rangle_{u+d}$ is truly extraordinary, displaying
a {\em minimum} for effective pion masses near 300-400 MeV before rising again to larger values for large quark masses.
  
As a second observation, we would like to note that the value for the generalized form
factor $B_{2,0}^s(t=0)$ could take on a {\em small negative} value at the physical point according to the right panel of Fig.\ref{Plotab}, 
albeit with a large uncertainty due to the 
poor data situation. Because of the small (negative!) value of the isoscalar Pauli form factor $F_2^s(t=0)\equiv \kappa_s=-0.12$ n.m., 
it is somewhat expected
that the next-higher moment\footnote{We can set $\xi=0$ in Eq.(\ref{f2}) and Eq.(\ref{momentsB}) for such a comparison.} 
will yield a value close to zero. However, Fig.\ref{Plotab} now opens
the possibility that $B_{2,0}^s(t=0) \approx -0.06$ might be
as large as 50\% of its $F_2^s(t=0)$ analogue. It will be 
very interesting to observe whether this feature can be reproduced when the new data of QCDSF and LHPC \cite{latticenext} at small pion 
mass are 
analyzed with the help of our ${\cal O}(p^2)$ BChPT formulae
Eqs.(\ref{xs},\ref{B20seq}).\\
 Finally, we note again that the true range of applicability of HBChPT
 versus covariant BChPT (see e.g. the left panel of Fig.(\ref{Plotab})) can only be determined, once the stability
 of the employed couplings is guaranteed, see the similar discussion
for $\langle x\rangle_{u-d}$ in section \ref{t0}. A study of higher order
 effects is therefore essential also in this respect.

\begin{table}[tb]
\begin{center}
\begin{tabular}{|c|c|c||c|c|}
\hline
$a_{2,0}^s$ & $b_{2,0}^s$ & $c_9$ & $\delta_A^0,\,\delta_B^0$ & $\langle x\rangle_{u+d}^{phen.}(\mu=2$GeV$)$ \\
$0.527\pm0.007$ & $-0.103\pm0.016$ &  $0.147\pm 0.002$ & $0$ (fixed) & $0.538 \pm 0.012$ (fixed) \\
\hline
\end{tabular}
\caption{The values for the three tensor coupling constants entering
$A_{2,0}^s(t)$ and $B_{2,0}^s(t)$ at order $p^2$ as extracted from a
combined fit to the lattice points for $A_{2,0}^s(0)$ and $B_{2,0}^s(0)$
\cite{QCDSF_dat} shown in Fig.\ref{Plotab} and to the physical point of $A_{2,0}^s(t=0,m_\pi=0.14$ GeV$)\equiv\langle x\rangle_{u+d}^{phen.}$. 
Note that we have obtained a small negative value for 
$b_{2,0}^s$. The indicated errors are only statistical and do not reflect the (much larger) systematic uncertainties connected with a fit
to the (quenched) lattice data of ref.\cite{QCDSF_dat} which {\it e.g.} neglect all contributions from disconnected 
diagrams.} \label{isoscalarfit}
\end{center}
\end{table}

\subsection{The contribution of $u-$ and $d-$quarks to the spin of the nucleon}

In the past few years a lot of interest in generalized isoscalar form factors of the nucleon has focused on the values of $A_{2,0}^s,\,B_{2,0}^s$ 
at the point $t=0$, 
as one can determine the contribution of quarks to the total spin of the nucleon via these two structures \cite{Ji}: 
\begin{eqnarray}
J_{u+d}&=&\frac{1}{2}\big[A_{2,0}^s(t=0)+B_{2,0}^s(t=0)\big].
\end{eqnarray}
To ${\cal O}(p^2)$ in BChPT we find
\begin{eqnarray}
J_{u+d}&=&\frac{1}{2}\bigg\{a_{2,0}^s+b_{2,0}^s\,\frac{M_N(m_\pi)}{M_0}+\frac{a_{2,0}^s\m^2}{(4\pi\F)^2}\bigg[\frac{3\g^2\m}{\sqrt{4\M^2-\m^2}}
	\bigg(8-6\frac{\m^2}{\M^2}+\frac{\m^4}{\M^4}\bigg)\arccos\frac{\m}{2\M} \nonumber\\
	& &-3\g^2\bigg(3-\frac{\m^2}{\M^2}+\bigg(2-4\frac{\m^2}{\M^2}+\frac{\m^4}{\M^4}\bigg)\log\frac{\m}{\M}\bigg)\bigg]+4
	m_\pi^2\frac{c_9}{M_0^2}\bigg\} +{\cal O}(p^3). \label{J}
\end{eqnarray}
Note that despite the plethora of non-analytic quark-mass dependent terms contained in the ${\cal O}(p^2)$ BChPT result of Eq.(\ref{J}), the 
two chiral logarithms calculated in ref.\cite{CJ2} within the HBChPT framework are {\em not yet} contained in our result. Both terms 
($\sim a_{(q)\pi}, \, b_{(q)N}$ in the notation of ref.\cite{CJ2}) are part of the complete ${\cal O}(p^3)$ result according to our power-counting 
and will appear in the calculation of the next order\footnote{The
contribution $\sim a_{(q)\pi}$ is already contained in the function 
$\Delta B_{h.o}(t=0,m_\pi)$ discussed in subsection \ref{firstglance}.}. We further note that the two logarithms of ref.\cite{CJ2} are 
UV-divergent and are accompanied by
a counterterm, whereas the ${\cal O}(p^2)$ BChPT result of Eq.(\ref{J}) happens to be UV-finite to the order we are working in. In ref.\cite{CJ2}
the authors also reported that the two chiral logarithms (of ${\cal O}(p^3)$) which they describe presumably are canceled {\em numerically} 
by pion-cloud contributions around an intermediate Delta(1232) state. We can confirm that this possibility exists, as the described Delta
contributions also start at ${\cal O}(p^3)$, assuming a power-counting where the nucleon-Delta mass difference is counted as 
parameter $\sim p^1$ (in the chiral limit) (see ref.\cite{SSE} for details).
 
Utilizing the ${\cal O}(p^2)$ BChPT result of Eq.(\ref{J})  and the fit parameters of table \ref{isoscalarfit} we obtain a a first estimate for
the contribution of $u-$ and $d-$quarks to the spin of a nucleon:  
\begin{equation}
J_{u+d}(t=0,\,m_\pi=0.14\,\rm{GeV})\approx 0.24 \pm 0.05,
\end{equation} 
which is only 50\% of the total spin of the nucleon! Note that we can only give a rough estimate for the error in this determination via assigning a 
typical size of the input lattice error bars shown in Fig.\ref{Plotab} to the extrapolated result, as we are assuming that the true error is dominated by
{\em systematic} errors in the lattice input to our analysis. (The
size of the statistical error read off from the fit of table
\ref{isoscalarfit} is $\pm 0.01$  and therefore negligible. The small value of this statistical error is of course heavily influenced by the error
assigned to the phenomenological value of $\langle x\rangle_{u+d}$ given in table \ref{isoscalarfit})

\begin{figure}[tb]
  \begin{center}
   \includegraphics*[width=0.5\textwidth]{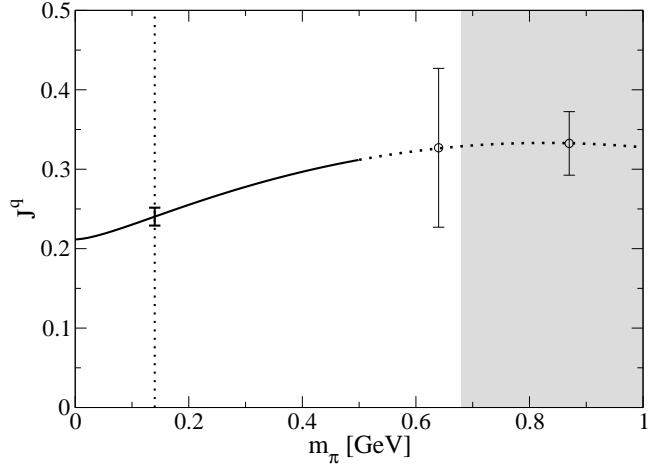}
    \caption{The contribution of $u+d$ quarks to the spin of a nucleon as function of the effective pion mass. The ${\cal O}(p^2)$ BChPT 
result shown as the solid line
is a prediction of Eq.(\ref{J}) which utilizes the fit-parameters of table \ref{isoscalarfit}. For comparison we have also plotted simulation data
from QCDSF \cite{QCDSF_dat} in the figure. At the physical point one can read off $J_{u+d}\approx 0.24$. (The error bar shown at the physical point is
only statistical and does not reflect any systematic uncertainties.)}
\label{PlotA+B}
  \end{center}
\end{figure}

Based on the same input (uncertainties) we can also predict the quark-mass dependence of $J_{u+d}$. The result is displayed 
as the solid line in Fig.\ref{PlotA+B}. 
Note that in contrast to the analysis given in ref.\cite{QCDSF_dat}, we do not obtain a flat chiral extrapolation function between 
the lattice data and the physical point.
The ${\cal O}(p^2)$ BChPT analysis suggests that the value at the physical point lies {\em lower} than the values 
obtained in the QCDSF simulation at large quark-masses.
However, given the large values of quark-masses and the sizable error-bars in the (quenched) 
simulation of ref.\cite{QCDSF_dat}, our extrapolated value for $J_{u+d}$
obviously can only give a first estimate of the true result. We are therefore looking forward to the application of our 
formula Eq.(\ref{J}) to the new (fully dynamical) 
lattice QCD results by QCDSF and LHPC obtained at lower quark-masses with improved statistics \cite{latticenext}. Furthermore, the 
influence of possible corrections from ${\cal O}(p^3)$ in Eq.(\ref{J}) has to be analyzed, before one can obtain a high precision
determination of the quark-contribution to the spin of the nucleon from a combined effort of lattice QCD and ChEFT \cite{future}.

\subsection{A first glance at the generalized isoscalar form factors of the nucleon}\label{firstglance}

In this section we present the results for the momentum  and
quark-mass dependence of the generalized form factors of the 
nucleon for the isoscalar flavour combination 
$u+d$ at ${\cal O}(p^2)$ in BChPT. We note that at this order the only non-zero loop contributions to the three isoscalar form factors arise
 from diagrams c) and e) in Fig.\ref{diags}, as the coupling of the isoscalar tensor field to the nucleon is not affected by chiral rotations. One 
obtains
\begin{eqnarray}
A_{2,0}^s(t) & = & A_{2,0}^s(0)-\frac{a_{2,0}^s g_A^2}{64 \pi^2 F_{\pi}^2} F_{2,0}^s(t)+\frac{c_{13}}{M_0^2}\,t+\mathcal{O}(p^3)\,, \label{A20s}
\end{eqnarray}
with $A_{2,0}^s(0)$ given in Eq.(\ref{xs}) and $F_{2,0}^s(t=0)\equiv 0$.
Interestingly, to the order we are working here, the $t$-dependence of this isoscalar form factor $A_{2,0}^s(t)$ is given by the same function 
\begin{eqnarray}
F_{2,0}^s(t)& = & F_{2,0}^v(t) +{\cal O}(p^3),
\end{eqnarray}
that controls its isovector analogue Eq.(\ref{davt}), albeit with {\em larger} numerical prefactors (compare Eq.(\ref{A20s}) to Eq.(\ref{A20v})).
We note that $F_{2,0}^s(t)$ does not depend on the scale $\lambda$ of dimensional regularization for the loop diagrams. The chiral coupling
$c_{13}$ is therefore also scale-independent\footnote{We note again, that this scale-independence refers to the UV-scales of the ChEFT 
calculation, not to be confused with the scale- and scheme-dependence of the quark-operators on the left hand side of Eq.(\ref{basic}), which is
completely outside the framework of ChEFT.}, parametrizing the (quark-mass independent) short-distance contributions to the 
radius/slope of $A_{2,0}^s(t)$. 
The unknown contributions from higher orders in the chiral expansion can be estimated from a calculation of the triangle diagram displayed in
Fig.\ref{tensorpi}. Due to the coupling of the tensor field to the (long-range) pion cloud of the nucleon, among all the contributions at the next 
chiral order this diagram should give the most-important $t$-dependent correction to the covariant ${\cal O}(p^2)$ result of Eq.(\ref{A20s}), 
resulting in the estimate ${\cal O}(p^3)\sim\Delta A^s_{h.o.}(t,m_\pi)$. The explicit expression for this function can be found in appendix 
\ref{sec:isosp3}. For completeness, we also note that in the limit $1/(16\pi^2F_\pi^2M_0)\rightarrow 0$ we obtain the corresponding 
${\cal O}(p^2)$ HBChPT result
\begin{eqnarray}
A_{2,0}^s(t)|^{p^2}_{HBChPT}&=&a_{2,0}^s+4m_{\pi}^2\frac{c_9}{M_0^2}+\frac{c_{13}}{M_0^2}\,t+{\cal O}(1/(16\pi^2F_\pi^2M_0))\, ,
\end{eqnarray}
which is just a string of tree level couplings.

To order ${\cal O}(p^2)$ in BChPT the two other isoscalar form factors
read

\begin{eqnarray}
B_{2,0}^s(t) & = & b_{2,0}^s\,\frac{\Mn(m_{\pi})}{\M}-\frac{a_{2,0}^s
	g_A^2M_0^2}{16\pi^2F_{\pi}^2}\int_{-\frac{1}{2}}^{\frac{1}{2}}\!
	\frac{du}{\tilde{M}^8}\Bigg\{\left(M_0^2-\tilde{M}^2\right)\tilde{M}^6+9m_{\pi}^2M_0^2\tilde{M}^4\nonumber \\ && -6m_{\pi}^4M_0^2\tilde{M}^2+6m_{\pi}^2M_0^2
	\left(m_{\pi}^4-3m_{\pi}^2\tilde{M}^2+\tilde{M}^4\right)\log{\frac{m_{\pi}}{\tilde{M}}}\nonumber \\ &&-\frac{6m_{\pi}^3M_0^2}{\sqrt{4\tilde{M}^2-m_{\pi}^2}}\bigg[
	m_{\pi}^4-5m_{\pi}^2\tilde{M}^2+5\tilde{M}^4\bigg]\arccos{\left(\frac{m_{\pi}}{2\tilde{M}}\right)}\Bigg\}\nonumber \\ &&
	+\Delta B_{h.o.}^s(t,m_{\pi}),\label{B20s}
\end{eqnarray}
\begin{eqnarray}
C_{2,0}^s(t) & = & c_{2,0}^s\frac{M_N(m_\pi)}{M_0}-
	\frac{a_{2,0}^sg_A^2M_0^2}{16\pi^2F_{\pi}^2}\int_{-\frac{1}{2}}^{\frac{1}{2}}\frac{du\,u^2}{\tilde{M}^8}\Bigg\{2\left(M_0^2-\tilde{M}^2\right)\tilde{M}^6
	-3m_{\pi}^2M_0^2\tilde{M}^4\nonumber \\&&+6m_{\pi}^4M_0^2\tilde{M}^2-6m_{\pi}^4M_0^2\left(m_{\pi}^2-2\tilde{M}^2\right)\log{\frac{m_{\pi}}{\tilde{M}}}
	\nonumber
	\\&&+\frac{6m_{\pi}^3M_0^2}{\sqrt{4\tilde{M}^2-m_{\pi}^2}}\bigg[m_{\pi}^4-4m_{\pi}^2\tilde{M}^2+2\tilde{M}^4\bigg]\arccos{
	\left(\frac{m_{\pi}}{2\tilde{M}}\right)}\Bigg\}
	\nonumber \\&&+\Delta C_{h.o.}^s(t,m_{\pi}).\label{C20s}
\end{eqnarray}
with $\tilde{M}$ defined in appendix \ref{sec:amplitudes} (c.f. Eq.(\ref{tM})).
$M_N(m_\pi)$ denotes again the (quark-mass dependent) mass function of the nucleon Eq.(\ref{p4mass}), introduced via Eq.(\ref{defs}). 
In the limit $1/(16\pi^2F_\pi^2M_0)\rightarrow 0$ we obtain the corresponding ${\cal O}(p^2)$ HBChPT results for 
$B_{2,0}^s(t),\, C_{2,0}^s(t)$, which at this order only consist of the tree-level couplings $b_{2,0}^s,\,c_{2,0}^s$. As in the case of $A_{2,0}^s(t)$
we have estimated the contributions from higher orders via ${\cal O}(p^3)\sim\Delta B_{h.o}^s(t,m_\pi),\,\Delta C_{h.o}^s(t,m_\pi)$, assuming
that the dominant $t$-dependent higher order corrections to our covariant ${\cal O}(p^2)$ BChPT results of Eqs.(\ref{B20s},\ref{C20s}) 
originate
from the ${\cal O}(p^3)$ triangle diagram displayed in Fig.\ref{tensorpi}. Explicit expressions are given in appendix \ref{sec:isosp3}. We 
note that the non-analytic 
quark-mass dependent terms in $A_{2,0}^s(t),\,B_{2,0}^s(t),\, C_{2,0}^s(t)$ calculated in ref.\cite{BJ} with the help of the the HBChPT
formalism\footnote{In refs.\cite{ACK,DM} additional terms have been calculated within the
HBChPT approach. While some terms correspond to ${\cal O}(p^3)$ and ${\cal O}(p^4)$ contributions according to our 
power-counting, expanding the covariant ${\cal O}(p^2)$ result of Eq.(\ref{B20s}) to the order 
$\left(\frac{1}{16\pi^2F_\pi^2M_0}\right)^1$  one can {\it e.g.} also
recognize a term $\sim a_{2,0}^s\,m_\pi^2$ present in ref.\cite{ACK}. However, as far
as we can see, neither ref.\cite{ACK} nor ref.\cite{DM} presents a {\em complete} ${\cal O}(p^3)$ HBChPT calculation of the 
{\em matrix element} 
Eq.(\ref{defs}).}  correspond to the leading terms in a $1/(16\pi^2F_\pi^2M_0)$ truncation of the ${\cal O}(p^3)$ BChPT corrections 
$\Delta X_{h.o.}(t,m_\pi),\;X=A,B,C$ of appendix \ref{sec:isosp3}. 

At this point we refrain from a detailed numerical analysis of the $t$-dependence of the generalized isoscalar form factors 
$A_{2,0}^s(t),\,B_{2,0}^s(t)$.
On the one hand very few lattice data for this flavour 
combination have been published so far for pion masses below 600 MeV. Moreover, available 
lattice data neglect contributions from ``disconnected diagrams'' and are therefore accompanied by an unknown systematic uncertainty, which 
is very hard to estimate. On the other hand, in the $t$-dependence both of $A_{2,0}^s(t)$ and of $B_{2,0}^s(t)$ we encounter chiral couplings 
($c_{13}$ at ${\cal O}(p^2)$ in Eq.(\ref{A20s}) and $B_{34}$ at ${\cal O}(p^3)$ in $\Delta B_{h.o.}^s(t,m_\pi)$ of Eq.(\ref{db})) connected
with (presently unknown) short-distance physics. The resulting leading $t$-dependence of these two generalized isoscalar form factors is 
hard to quantify at this point without any additional input either from experiment or from lattice QCD. We are therefore postponing this
discussion until further information is available, {\it e.g.} from the new lattice QCD simulations at low momentum transfer \cite{latticenext}.
In the meantime we are preparing \cite{future} a full (next-to-leading one loop) ${\cal O}(p^3)$ BChPT analysis of the 
isoscalar moments of the GPDs which (in addition to several other diagrams!) also contains the contributions from the triangle diagram 
shown in Fig.\ref{tensorpi}, already presented in appendix \ref{sec:isosp3}. 

Before finally proceeding to the summary of this work, we want to take a look at the third generalized isoscalar form factor $C_{2,0}^s(t)$ of
Eq.(\ref{C20s}). According to our power-counting, short distance contributions to the radius of this form factor are suppressed and
only start to enter at ${\cal O}(p^4)$, both in HBChPT and in BChPT. After adding the ${\cal O}(p^3)$ {\em estimate} 
$\Delta C_{h.o.}^s(t,m_\pi)$ to the ${\cal O}(p^2)$ BChPT result of Eq.(\ref{C20s}), we can hope to catch a first glance of the $t$-dependence
of this elusive nucleon structure. Utilizing $a_{2,0}^s$ of table \ref{isoscalarfit} and assuming $x_\pi^0\approx \langle x\rangle_\pi^s\approx 0.5$ 
at a renormalization scale $\mu^2=4$ GeV$^2$ \cite{DrellYang} we can determine its slope  
\begin{eqnarray}
\rho_C^s & = & \frac{d\,C_{2,0}^s(t)}{d\, t}|_{t=0} \nonumber \\
                & =&\frac{g_A^2}{640 \pi^2
	F_{\pi}^2M_0^6}\Bigg\{4a_{2,0}^sm_{\pi}^4\left(2m_{\pi}^2-3M_0^2\right)\log{\frac{m_{\pi}}{M_0}}+a_{2,0}^s\bigg(
	m_{\pi}^6-8m_{\pi}^4M_0^2+2m_{\pi}^2M_0^4-\frac{2}{3}M_0^6\bigg)\nonumber \\&& 
	+x_{\pi}^0M_0 M_N(m_{\pi})\bigg(7m_{\pi}^4-27 m_{\pi}^2M_0^2+\frac{20}{3}M_0^4\bigg)-2x_{\pi}^0\frac{M_N(m_{\pi})}{M_0}
	\bigg(4m_{\pi}^6-21m_{\pi}^4M_0^2\nonumber \\&& +20m_{\pi}^2M_0^4+5M_0^6\bigg)\log{\frac{m_{\pi}}{M_0}}
	-\frac{1}{m_{\pi}^3-4m_{\pi}M_0^2}\Bigg[
	a_{2,0}^s\bigg(m_{\pi}^8-4M_0^2m_{\pi}^6+2M_0^4m_{\pi}^4\bigg)\nonumber \\&&-x_{\pi}^0 M_0M_N(m_{\pi})\bigg(m_{\pi}^6-7m_{\pi}^4M_0^2+9m_{\pi}^2M_0^4+8M_0^6\bigg)
	-\frac{1}{\sqrt{4M_0^2-m_{\pi}^2}}\Bigg(4a_{2,0}^sm_{\pi}^4\left(-2m_{\pi}^6\right.\nonumber \\ && \left. +15m_{\pi}^4M_0^2-30m_{\pi}^2M_0^4+10M_0^6\right)
	+2x_{\pi}^0\frac{M_N(m_{\pi})}{M_0}\left(
	4m_{\pi}^{10}-45M_0^2m_{\pi}^8+170m_{\pi}^6M_0^4\right.\nonumber \\&& \left. -225m_{\pi}^4M_0^6+30 m_{\pi}^2M_0^8+32M_0^{10}\right)\Bigg)\Bigg]
            \arccos{\left(\frac{m_{\pi}}{2M_0}\right)}\Bigg\}. \nonumber \\
& &\label{rc}
\end{eqnarray}
At the physical point this would give us $\rho_C^s(m_{\pi}=0.14 $GeV$) = -0.77$ GeV$^{-2}$. 
Truncating Eq.(\ref{rc}) in $1/(16\pi^2F_\pi^2M_0)$ we reproduce the chiral singularity $\sim m_\pi^{-1}$ found in 
ref.\cite{BJ}\footnote{Note that due to a different definition of the covariant derivative in the quark-operator on the 
left hand side of Eq.(\ref{basic}) our
definition of the third isoscalar form factor differs from ref.\cite{BJ} by a factor of 4: $C_2^{BJ}(t)\equiv 4 C_{2,0}^s(t)$.} 
\begin{equation}
\rho_C^s =  -\frac{g_A^2x_{\pi}^0 M_N(m_{\pi})}{160 \pi
	F_{\pi}^2 m_{\pi}}-\frac{g_A^2}{960  \pi^2
	F_{\pi}^2}\left[a_{2,0}^s+x_{\pi}^0\frac{M_N(m_{\pi})}{M_0}\left(-13+15\log{\frac{m_{\pi}}{M_0}}\right)\right]+
	... \label{rcchi}
\end{equation}
It amounts to a slope of $\rho_C^s|_\chi=-1.12$ GeV$^{-2}$, which is 45\% larger than the BChPT estimate of Eq.(\ref{rc}).
Interestingly, among the terms $\sim m_\pi^0$ shown in Eq.(\ref{rcchi}) it is the $1/M_0$ suppressed corrections to the leading HBChPT 
result of ref.\cite{BJ} that dominate numerically. This gives a strong indication that a {\em covariant} calculation of $\Delta C_{h.o.}^s(t,m_\pi)$ 
as given in appendix \ref{sec:isosp3} is advisable, automatically containing {\em all} associated $(1/M_0)^n$ corrections.

\begin{figure}[tb]
  \begin{center}
   \includegraphics*[width=0.5\textwidth]{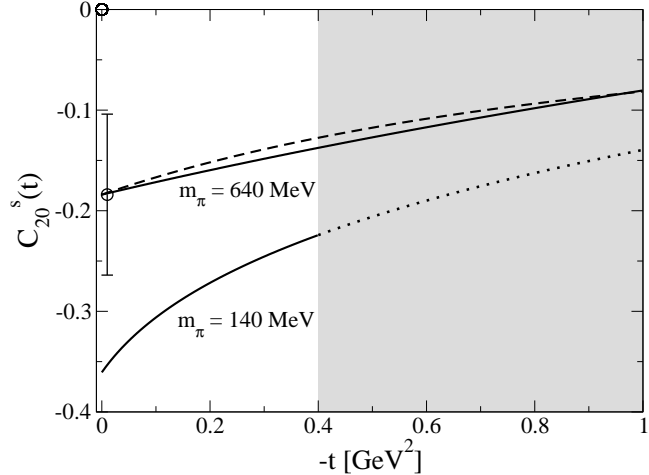}
    \caption{A first glance at the momentum dependence of the form
factor $C_{2,0}^s(t)$. The quasi-linear $t-$dependence at the  pion-mass 
$m_\pi=640$ MeV (solid line) has been normalized to the dipole parametrization of
the QCDSF data of ref.\cite{QCDSF_dat} (dashed line). The resulting non-linear $t$-dependence in this
form factor for smaller values of $m_\pi$ is then due to the coupling of the tensor field to the pion cloud, providing an interesting mechanism
to obtain ``large'' negative values at $t=0,\,m_\pi=140$ MeV. } 
\label{Plotcst}
  \end{center}
\end{figure}

The quark-mass dependence of the slope function $\rho_C^s$ of Eq.(\ref{rc}) already suggests that one obtains an interesting variation of the $t$-dependence 
of this form factor as a function of the quark-mass! We therefore close this discussion with a look at Fig.\ref{Plotcst}. There we have fixed
the only unknown parameter $c_{2,0}^s=-0.41\pm 0.1$ such that the BChPT result
coincides with the dipole parametrization of the  QCDSF collaboration at $t=0$
\cite{QCDSF_dat} for the lightest pion mass in the simulation, i.e. $m_{\pi}=640$ MeV. We note explicitly that this coupling only affects the overall normalization 
of this form factor, but does not impact its momentum-dependence. It is therefore quite remarkable to observe that  the resulting $t-$dependence of this form factor 
according to this ChEFT estimate agrees quite well with the phenomenological dipole-parametrization of the QCDSF data at this large quark-mass, even over 
quite a long range in four-momentum 
transfer. The result is rather close to a straight line\footnote{Most of the pion loop diagrams in BChPT with IR regularization as 
employed in appendix \ref{sec:RF} have the remarkable 
feature that they automatically go to zero in the limit of large pion masses, without any extra assumptions. This aspect of BChPT will be 
discussed in detail in ref.\cite{GH}.} (see Fig.\ref{Plotcst}). We remind the reader that the value of this form factor at $t=0$ and $m_\pi=0.14$ GeV determines
the strength of the so-called D-term of the nucleon, playing a decisive role in the analysis of Deeply Virtual Compton Scattering (DVCS) 
experiments \cite{reviews}. Utilizing the extracted value of $c_{2,0}^s$ we can now study the C-form factor also at the physical 
point, with the result also shown in Fig.\ref{Plotcst}. At this low value of the pion mass one can suddenly observe a non-linear $t-$dependence
 for low values of four-momentum transfer, {\em due to the pion cloud of the nucleon}.
This is a very interesting observation, because such a mechanism would allow for a more negative value of $C_{2,0}^s(t=0)$ at the 
physical point than previously extracted from lattice QCD analyses via dipole extrapolations ({\it e.g.} see ref.\cite{QCDSF_dat}). We obtain
\begin{equation}
C_{2,0}^s\left(t=0,m_\pi=140\,\rm{MeV}\right)\approx\,-0.36 \pm 0.1,
\end{equation}
The assigned error corresponds to the fit error of $C_{2,0}^s(m_\pi=0.64\,\rm{GeV},t=0)_{QCDSF}$ given in ref.\cite{QCDSF_dat}, as it directly influences our unknown 
coupling $c_{2,0}^s$. However, we note that the (unknown) systematic uncertainties underlying the (quenched) simulation results of ref.\cite{QCDSF_dat} are not 
accounted for in this error bar. A comparison of these data with the new dynamical simulations of QCDSF and LHPC \cite{latticenext} might shed more light onto this unresolved
question of systematic uncertainties in lattice QCD\footnote{However, according to our understanding even the new dynamical simulation results of QCDSF and LHPC will 
neglect all contributions from disconnected diagrams.}. 

With this value we can finally obtain the first estimate for the radius of this elusive form factor:
\begin{eqnarray}
\left(r_C^s\right)^2&=&\frac{6}{C_{2,0}^s(t=0,\,m_\pi=140\,\rm{MeV})}\,\rho_C^s(m_\pi=140\,\rm{MeV}) \nonumber \\
&\approx&(0.5\pm 0.1)\, \rm{fm}^2. 
\end{eqnarray}
We compare this result with the radii of the isovector Dirac and Pauli form factors of the nucleon, which are also dominated by pion cloud
effects. Interestingly, with $\left(r_1^v\right)^2= 0.585$ fm$^2$ and $\left(r_2^v\right)^2=0.80$ fm$^2$ \cite{disp} 
the estimated value for $\left(r_C^s\right)^2$
seems to lie in the same order of magnitude! We note, however, that our numerical estimate of its slope $\rho_C^s$ of Eq.(\ref{rc}) given
above is {\em significantly} smaller than the corresponding 
slopes of the isovector Pauli and Dirac form factors, as expected from general arguments and as already observed in lattice 
QCD simulations with dynamical fermions \cite{LHPC-SESAM} (at very large quark masses).

However, before we can go into a more detailed numerical
analysis of these interesting new form factors of the nucleon, one should first complete the ${\cal O}(p^3)$ calculation of the generalized 
isoscalar form factors,
as there are additional diagrams next to Fig.\ref{tensorpi} possibly also affecting the $t$-dependence, albeit presumably in a weaker fashion 
\cite{future}. 

\section{Summary} 
\label{sec:summary} 
The pertinent results of this analysis can be summarized as follows:
\begin{enumerate}
\item We have constructed the effective chiral Lagrangean for symmetric, traceless tensor fields of positive parity up to ${\cal O}(p^2)$ in
the covariant framework of Baryon Chiral Perturbation Theory for 2 light quark flavours. 
\item Within this covariant framework we have calculated the generalized isovector and isoscalar form factors of the nucleon 
$A_{2,0}^{v,s}(t,m_\pi^2),\, B_{2,0}^{v,s}(t,m_\pi^2),\, C_{2,0}^{v,s}(t,m_\pi^2)$ up to ${\cal O}(p^2)$, which 
corresponds to leading-one-loop order. We can exactly reproduce the corresponding non-relativistic ${\cal O}(p^2)$ results previously 
obtained in Heavy Baryon ChPT by taking the limit $1/(16\pi^2F_\pi^2M_0)\rightarrow 0$. Several HBChPT results published recently
could not (yet) be reproduced, as they correspond to partial, non-relativistic results from the higher orders ${\cal O}(p^3,p^4,p^5)$.
\item According to our numerical analysis of the quark-mass dependence of the generalized form factors, we have noted that for 
$B_{2,0}^{s,v}(t)$ and  $C_{2,0}^{s,v}(t)$
the observable quark-mass dependencies could be dominated by the (well-known) quark-mass
 dependence of the mass of the nucleon $M_N(m_\pi^2)$. This mass function appears in several places in the chiral results due to 
kinematical factors in the matrix-element used in the definition of the
generalized form factors. Such a ``trivial'' but numerically significant effect is already known from the analysis of lattice QCD data for the 
Pauli form factors of the nucleon. 
\item The pion-cloud contributions to all three generalized {\em isovector} form factors at finite values of $t$ are very small. The momentum 
dependence 
of these structures seems to be dominated by (presently) unknown short distance contributions. The situation in this isovector channel 
reminds us of an analogous role played by chiral dynamics in the
isoscalar Dirac and Pauli form factors of the nucleon. At this point we were therefore not able to give predictions for the {\em numerical} size 
of the radii or slopes
of these interesting nucleon structure quantities. It is hoped that a global fit to new lattice QCD data at small pion masses and small values of
$t$---extrapolated to the physical point with the help of the formulae presented in this work---will lead to first insights into this new field of 
baryon structure physics \cite{latticenext}.
\item The BChPT leading-one-loop results for the  generalized {\em isoscalar} form factors $A_{2,0}^s(t),\,B_{2,0}^s(t)$ are quite surprising. 
As far as the topology of possible Feynman diagrams is 
concerned, one is reminded of the isovector Dirac and Pauli form factors of the nucleon. A power-counting analysis, however, told us that those
diagrams ({\it e.g} see Fig.\ref{tensorpi}) from which one would expect large non-analytic contributions in the quark-mass 
only start to contribute at next-to-leading one-loop order.
Our analysis therefore suggests that the momentum-dependence at low values of $t$ is dominated by 
short-distance physics. A new quality of lattice QCD data for low pion masses and low values
of $t$ to constrain some of the unknown couplings is urgently needed before further conclusions can be drawn \cite{latticenext}.
\item It is the 
value of $\langle x\rangle_\pi$ of a pion in the chiral limit that controls the magnitude of those long-distance pion-cloud effects in the
generalized isoscalar form factors of the nucleon, pointing to the
 need of a simultaneous analysis of pion and nucleon structure on the lattice and in ChEFT.   
\item In the forward limit, the isovector form factor $A_{2,0}^v(t\rightarrow 0)$ reduces to $\langle x \rangle_{u-d}$. 
Our covariant ${\cal O}(p^2)$ BChPT
result for this isovector moment provides a smooth chiral extrapolation function between the high values at large quark-masses from the 
LHPC collaboration and the lower value known from phenomenology. The required (chiral) curvature according to this new analysis 
does {\em not} originate from the 
chiral logarithm of the leading-non-analytic quark-mass dependence of this moment---as had been speculated in the literature for the past few
years---but is due to an infinite tower of terms $(m_\pi/M_0)^n$ with well-constrained coefficients (see Eq.(\ref{A20v0})). 
The well-known leading-one-loop HBChPT result for $\langle x
\rangle_{u-d}$ of Eq.(\ref{xu-dHBChPT}) was found to be applicable only
for chiral extrapolations from the chiral limit to values slightly above the physical pion mass, as expected. 
\item Judging from the available (quenched) lattice data of the QCDSF collaboration
our ${\cal O}(p^2)$ BChPT result of Eq.(\ref{xs}) for $A_{2,0}^s(t=0)\equiv\langle x\rangle_{u+d}$ also provides a very stable chiral 
extrapolation function out to quite large values of effective pion masses. 
\item A study of the forward limit in the isoscalar sector has led to a first estimate of the contribution of the $u-$ and $d-$quarks to the total
spin of a nucleon $J_{u+d}\approx 0.24$. This low value compared to previous determinations arises from the possibility of a small negative
contribution of $B_{2,0}^s(t=0)\approx -0.06$ at the physical point, driven by pion cloud effects. However, at the moment the uncertainty in such a 
determination is rather large, due to the poor situation of available data from lattice QCD. 
\item  In a first glance at the third generalized isoscalar form factor $C_{2,0}^s(t)$ the quark-mass dependence was found to be qualitatively 
different from $A_{2,0}^s(t),\,B_{2,0}^s(t)$. Its slope contains a chiral singularity  $\sim m_\pi^{-1}$ and the influence of short distance 
contributions is suppressed. A first 
numerical estimate of its slope gives $\rho_C^s\approx-0.75$ GeV$^2$, which is much smaller than the slopes of corresponding
Dirac or Pauli form factors. At low t we have also observed significant changes in the momentum 
dependence of this form factor as a function of the quark-mass, resulting in the estimate $C_{2,0}^s(t=0)\approx -0.35$ at the 
physical point. 
\item Throughout this work we have indicated how to estimate possible corrections of higher orders to our leading-one-loop BChPT results. The 
associated theoretical uncertainties of our ${\cal O}(p^2)$ calculation have been discussed in detail. Ultimately, in order to judge the stability 
of our results it is mandatory that we analyse the complete next-to-leading one-loop order. Work in this direction has already started \cite{future}.
\end{enumerate}
Finally, we note that the tensor Lagrangeans constructed in section \ref{sec:formalism} invite a host of further studies, pertaining both to 
generalized
axial form factors of the nucleon \cite{DHaxial} and to the energy-momentum-tensor of the nucleon \cite{EMT}. 
 
\bigskip

\section*{Acknowledgements}
MD would like to acknowledge financial support from the University of Pavia and is grateful for the hospitality and financial support of the group
T39 at TU M{\" u}nchen. The work of TAG and TRH has been supported under the EU Integrated Infrastructure Initiative Hadron Physics under
contract number RII3-CT-2004-506078. The authors are indebted to Ph. H{\" a}gler for critical advice regarding the available data base
of lattice QCD studies on nucleon GPDs and also acknowledge helpful
discussions with B. Musch, B. Pasquini, M. Procura, G. Schierholz and W. Weise.
The authors are also grateful to the LHPC collaboration for providing their (preliminary) lattice data on 
$\langle x\rangle_{u-d}$ of ref.\cite{LHPC}. The diagrams in the paper have been drawn using the Jaxodraw package \cite{BT}.

\newpage

\appendix

\section{Basic Integrals}
\def\theequation{\Alph{section}.\arabic{equation}}
\setcounter{equation}{0}
\label{sec:integrals}

The integrals required for one-loop calculations in BChPT can be reduced to two basis integrals in d-dimensions:
\begin{eqnarray}
\Delta_\pi\left(m\right) &\equiv& \frac{1}{i}\int\frac{d^dl}{(2\pi)^d}\;\frac{1}{m^2-l^2-i\epsilon}, \label{b1}\\
H_{11}\left(M^2,m^2,p^2\right)&\equiv&\frac{1}{i}\int\frac{d^dl}{(2\pi)^d}\;\frac{1}{\left(m^2-l^2-i\epsilon\right)\left(M^2-(l-p)^2-
i\epsilon\right)}, \label{b3} 
\end{eqnarray}
where $m\; (M)$ is a mass function involving the mass of the Goldstone Boson (of the Baryon) and $p^\mu$ denotes a four-momentum 
determined by the kinematics.The propagators are shifted into the complex energy-plane by a small amount $\epsilon$ to ensure causality. 
Utilizing the $\overline{MS}$-renormalization scheme of ref.\cite{GSS} with 
\begin{equation}
L=\frac{\lambda^{d-4}}{16\pi^2}\left[\frac{1}{d-4}+\frac{1}{2}\left(\gamma_E-1-\ln 4\pi\right)\right],
\end{equation} 
one obtains the dimensionally-regularized results \cite{GSS}
\begin{eqnarray}
\Delta_{\pi} &=& 2m^2\bigg(L+\frac{1}{16\pi^2} \ln\frac{m}{\lambda}\bigg)+\mathcal{O}(d-4),\\
H_{11}(M^2,m^2,p^2) &=& -2L-\frac{1}{16
\pi^2}\bigg[-1+\log\frac{M^2}{\lambda^2}+\frac{p^2-M^2+m^2}{p^2}\log\frac{m}{M}
\nonumber  \\ &&
+\frac{2mM}{p^2}\sqrt{1-\bigg(\frac{p^2-M^2-m^2}{2mM}\bigg)^2}\arccos\bigg(\frac{m^2+M^2-p^2}{2mM}\bigg)\bigg] \nonumber \\
& &+\mathcal{O}(d-4).
\end{eqnarray}
More complicated integral expressions needed during the calculation are defined via
\begin{align}
\frac{1}{i}\int\frac{d^dl}
{(2\pi)^d}\,\frac{\{l^{\mu},l^{\mu}l^{\nu}\}}{(m^2-l^2-i\epsilon)(M^2-(l-p)^2-i\epsilon)}=&\{p^\mu H_{11}^{(1)},g^{\mu\nu}H_{11}^{(2)}
+p^{\mu}p^{\nu}H_{11}^{(3)}\}, \label{h1} \\
\frac{1}{i}\int\frac{d^dl}
{(2\pi)^d}\,\frac{l^{\mu}l^{\nu}l^{\alpha}}{(m^2-l^2-i\epsilon)(M^2-(l-p)^2-i\epsilon)}=&\:(p^{\mu}g^{\mu\alpha}+p^{\nu}g^{\mu\alpha}+
p^{\alpha}g^{\mu\nu})\,H_{11}^{(4)}+p^{\mu}p^{\nu}p^{\alpha}\, H_{11}^{(5)}   \label{h2}\; .
\end{align}
The integrals $H_{11}^{(i)}$ are related to the two basis integrals of Eqs.(\ref{b1}-\ref{b3}) via well-known tensor-identities \cite{GSS,GH}. 
Finally, we note that the integrals involving more than one baryon propagator can be related to the ones defined above via
\begin{equation}
H_{11}^{(i)\prime}(M^2, m^2, p^2)\equiv\frac{\partial}{\partial M^2}H_{11}^{(i)}(M^2, m^2, p^2).
\end{equation}

\newpage
\section{Regulator Functions}
\def\theequation{\Alph{section}.\arabic{equation}}
\setcounter{equation}{0}
\label{sec:RF}

The basis regulator function is defined as \cite{BL,GH}
\begin{eqnarray}
R_{11}\left(M^2,m^2,p^2\right) & \equiv & \int_{x=1}^{\infty}\!dx\!\int\!\frac{d^dl}{(2\pi)^d}\left[xM^2+(x^2-x)p^2+(1-x)m^2-l^2\right]^{-2}.
\end{eqnarray}
More complicated regulator functions $R_{11}^{(i)},\,i=1\ldots 5$ can be defined in analogy to Eqs.(\ref{h1},\ref{h2}).
We note that for our ${\cal O}(p^2)$ calculation of the moments of the GPDs we only need to know these functions up to the 
power\footnote{Strictly speaking we need to know the regulator terms contributing to the generalized form factor $A_{2,0}^{s,v}(t)$ up to power $(m_\pi^2,t)^1$, 
whereas for $B_{2,0}^{s,v}(t), \,C_{2,0}^{s,v}(t)$ only the leading terms $(m_\pi^2,t)^0$ are required \cite{GH}.}  of $m_\pi^2,\,t$ in order
 to obtain a properly renormalized, scale-independent result, which at the same time is also consistent with the requirements of
 power-counting. The regulator functions needed for our ${\cal O}(p^2)$ BChPT calculation (see Appendix \ref{sec:amplitudes}) read
\begin{eqnarray}
R_{11} (M_0^2,m_{\pi}^2,p^2) & = &
	\left(2-\frac{m_{\pi}^2}{M_0^2}\right)L+\frac{1}{16\pi^2}\Bigg[2\log{\frac{M_0}{\lambda}}-1-\frac{1}{2}
	\frac{m_{\pi}^2}{M_0^2}\left(2\log{\frac{M_0}{\lambda}}+3\right)\Bigg]+...,\\
R_{11}^{(1)}(M_0^2,m_{\pi}^2,p^2) & = &
	\left(1+\frac{m_{\pi}^2}{M_0^2}\right)L+\frac{1}{16\pi^2}\Bigg[\log{\frac{M_0}{\lambda}}+\frac{1}{2}\frac{m_{\pi}^2}{M_0^2}
	\left(2\log{\frac{M_0}{\lambda}}-1\right)\Bigg]+...,\\
R_{11}^{(2)}(M_0^2,m_{\pi}^2,p^2) & = &
	\left(\frac{1}{3}M_0^2+\frac{1}{2}m_{\pi}^2\right)L+\frac{1}{48
	\pi^2}\Bigg[\frac{M_0^2}{3}\left(3\log{\frac{M_0}{\lambda}}-1\right)+\frac{3}{2}m_{\pi}^2\left(
	\log{\frac{M_0}{\lambda}}-1\right)\Bigg]\nonumber \\&&+...,\\
R_{11}^{(3)}(M_0^2,m_{\pi}^2,p^2) & = & \frac{2}{3}L+\frac{1}{48
	\pi^2}\Bigg[2\log{\frac{M_0}{\lambda}}+\frac{1}{3}+\frac{3}{2}\frac{m_{\pi}^2}{M_0^2}\Bigg]+...,
\end{eqnarray}
The derivatives of the regulator functions needed for the calculation (see Appendix \ref{sec:amplitudes}) read
\begin{eqnarray}
{R_{11}^{(2)}}^{\prime}(\tilde{M}^2,m_{\pi}^2,\tilde{p}^2) & = &
	\frac{1}{2}\left(1+\frac{m_{\pi}^2}{M_0^2}\right)L+\frac{1}{32\pi^2}
 	\Bigg[\log{\frac{M_0}{\lambda}}+\frac{m_{\pi}^2}{2M_0^2}\left(2\log{\frac{M_0}{\lambda}}-1\right)\Bigg]-\frac{t}{384\pi^2M_0^2}\nonumber \\ &&+...,\\
{R_{11}^{(3)}}^{\prime}(\tilde{M}^2,m_{\pi}^2,\tilde{p}^2) & = &
	-\frac{m_{\pi}^2}{M_0^4}L+\frac{1}{16\pi^2M_0^2}
	\Bigg[\frac{1}{2}+\frac{m_{\pi}^2}{M_0^2}\left(-\log{\frac{M_0}{\lambda}}+1\right)\Bigg]+\frac{t}{192\pi^2M_0^4}+...,\\
{R_{11}^{(4)}}^{\prime}(\tilde{M}^2,m_{\pi}^2,\tilde{p}^2) & = &
	\frac{1}{3}L+\frac{1}{16\pi^2}\Bigg[\frac{1}{3}\log{\frac{M_0}{\lambda}}+\frac{1}{18}+\frac{m_{\pi}^2}{4M_0^2}\Bigg]
	-\frac{t}{576\pi^2M_0^2}+...,\\
{R_{11}^{(5)}}^{\prime}(\tilde{M}^2,m_{\pi}^2,\tilde{p}^2) & = & \frac{1}{16\pi^2M_0^2}\Bigg[\frac{1}{3}-\frac{m_{\pi}^2}{2M_0^2}\Bigg]+\frac{t}{288\pi^2M_0^4}+...
\end{eqnarray}
One can clearly observe that all contributions are polynomial in $m_\pi^2$ (and therefore polynomial in the quark-mass \cite{BL}) or polynomial in $t$ \cite{GH}, as expected. 
Their addition to the 
$\overline{MS}$-results therefore just amounts to a shift in the coupling constants \cite{BL,GH} of the effective field theory and does not affect the non-analytic
quark-mass dependencies, which are the scheme-independent signatures of chiral dynamics. 

\newpage
\section{Isovector Amplitudes in ${\cal O}(p^2)$ BChPT}
\def\theequation{\Alph{section}.\arabic{equation}}
\setcounter{equation}{0}
\label{sec:amplitudes}

The five ${\cal O}(p^2)$ amplitudes in the isovector channel corresponding to the five diagrams of Fig.\ref{diags} written in terms of 
the basic integrals 
\begin{equation}
I_{11}^{(i)}\left(M^2,m^2,p^2\right)= H_{11}^{(i)}\left(M^2,m^2,p^2\right) +R_{11}^{(i)} \left(M^2,m^2,p^2\right)\; , \quad i=0\ldots 5,
\end{equation}
of Appendices~\ref{sec:integrals} and \ref{sec:RF}
 read
\begin{align}
Amp^{a+b}=&-i\:\frac{\Delta a_{2,0}^v\:\:\g}{\F^2}\:\:\eta^{\dagger}\frac{\tau^a}{2}\eta\:\:\overline{u}(p')
\:\gamma_{\{\mu}\overline{p}_{\nu\}}\:u(p)\:\bigg[2\m^2\I(\M^2, \m^2, p^2)\\&\nonumber-\m^2\I^{(1)}(\M^2, \m^2, p^2)+2 \I^{(2)}(\M^2, \m^2, p^2)\,\bigg],
\\
Amp^c=&\,i\:\frac{a_{2,0}^v \g^2}{4\F^2}\:\:\eta^{\dagger}\frac{\tau^a}{2}\eta\:\:\overline{u}(p')\:\int^{\frac{1}{2}}_{-\frac{1}{2}}\, du\:\:\\&\nonumber
	\Bigg\{\:\gamma_{\{\mu}\overline{p}_{\nu\}}\:\Bigg[-\Delta_\pi+4\M^2\bigg(\I^{(1)}(\M^2, \m^2, p^2)-\I^{(3)}(\M^2, \m^2, p^2)\\&\nonumber
	+(2M_0^2-\tilde{M}^2)\left(\I^{(3)\prime}(\tilde{M}^2, \m^2,\tilde{p}^2)-
	\I^{(5)\prime}(\tilde{M}^2, \m^2, \tilde{p}^2)\right)\\&\nonumber+(d-2)\left(\I^{(4)\prime}(\tilde{M}^2, 
	\m^2, \tilde{p}^2)-\I^{(2)\prime}(\tilde{M}^2, \m^2,\tilde{p}^2)\right)
	\bigg)\Bigg]\\&\nonumber+\,i\,\Delta^{\alpha}
	\sigma_{\alpha\{\mu}\overline{p}_{\nu\}}\:4\M^3\bigg(\I^{(3)\prime}(\tilde{M}^2, \m^2, \tilde{p}^2)-\I^{(5)\prime}(\tilde{M}^2, \m^2, \tilde{p}^2)\bigg)\\
	&\nonumber-\Delta_{\{\mu}\Delta_{\nu\}}\bigg(8\M^3\,u^2\,\I^{(5)\prime}(\tilde{M}^2, \m^2, \tilde{p}^2)\bigg)\Bigg\}\:u(p),\\
Amp^{d}=&-\,i\:\frac{a_{2,0}^v}{\F^2}\:\:\eta^{\dagger}\frac{\tau^a}{2}\eta\:\:\overline{u}(p')\:\gamma_{\{\mu}\overline{p}_{\nu\}}\:u(p)\:\Delta_\pi,\\
Amp^{e}=&i\:a_{2,0}^v\:\:\eta^{\dagger}\frac{\tau^a}{2}\eta\:\:\overline{u}(p')
\:\gamma_{\{\mu}\overline{p}_{\nu\}}\:u(p)\:\mathcal{Z_N}.
\end{align}
Note that the various couplings and parameters are defined in section \ref{sec:formalism}. $\eta$ denotes the isospin doublet of proton and neutron. 
The variables in the integral functions are given as
\begin{eqnarray}
\tilde{p}^2 \,\, = \,\,
\tilde{M}^2&\equiv&\M^2+\left(u^2-\frac{1}{4}\right)t, \label{tM}
\end{eqnarray}
where $t=\Delta^2$ corresponds to the momentum transfer by the tensor fields.

$\mathcal{Z_N}$ denotes the Z-factor of the nucleon, calculated to the required ${\cal O}(p^3)$ accuracy in BChPT. It is obtained from
 the self-energy $\Sigma_N$ at this order via the prescription
\begin{equation}
\mathcal{Z_N}=1+\frac{\partial \Sigma_N}{\partial \slashed{p}}\bigg\vert_{\slashed{p}=\M} + {\cal O}(p^4), 
\end{equation}
with
\begin{equation}
\Sigma_N=\frac{3\g^2}{4\F^2}(\M+\slashed{p})\big[\m^2 I_{11}(M_0^2,m_{\pi}^2,p^2)+(\M-\slashed{p})\slashed{p}I_{11}^{(1)}(M_0^2,m_{\pi}^2,p^2)-\Delta_N\big] + {\cal O}(p^4). 
\end{equation}

\newpage

\section{BChPT Results in the Isoscalar Channel}
\def\theequation{\Alph{section}.\arabic{equation}}
\setcounter{equation}{0}
\label{app:isoscalar}

\subsection{Isoscalar Amplitudes in ${\cal O}(p^2)$ BChPT}

To ${\cal O}(p^2)$ in BChPT the results in the isoscalar channel are quite simple. The amplitudes corresponding to the Feynman diagrams of 
Fig.\ref{diags} can be be simply expressed in terms of results already obtained in the isovector channel discussed in the previous section \ref{sec:amplitudes}. They read
\begin{eqnarray}
\overline{Amp}_{a+b}&=&0+{\cal O}(p^3), \\
\overline{Amp}_{c}&=&-3\,\frac{a_{2,0}^s \eta^\dagger {\bf 1} \eta}{a_{2,0}^v \eta^\dagger \tau^a \eta}\,Amp^c +{\cal O}(p^3),  \\
\overline{Amp}_{d}&=&0+{\cal O}(p^3), \\
\overline{Amp}_{e}&=&\frac{a_{2,0}^s \eta^\dagger {\bf 1} \eta}{a_{2,0}^v \eta^\dagger \tau^a \eta}\,Amp^e +{\cal O}(p^3). 
\end{eqnarray}
Note that the various couplings and parameters are defined in section \ref{sec:formalism}.

\subsection{Estimate of ${\cal O}(p^3)$ contributions}\label{sec:isosp3}

The contributions from the (higher order) Feynman diagram shown in figure \ref{tensorpi} to the
generalized isoscalar form factors read
\begin{eqnarray}
\Delta A^s_{h.o.}(t,m_\pi) & = & \frac{g_A^2
	x_\pi^0}{32\pi^2
	F_{\pi^2}}\int_{-\frac{1}{2}}^{\frac{1}{2}}\!\frac{du}{\tilde{p}^8}\,
	\Bigg\{2\tilde{p}^8\left(3m_{\pi}^2+M_0^2\right)-t\tilde{p}^8+4M_0^2\left(2M_0^2+\tilde{m}^2\right)\tilde{p}^6
	\nonumber \\&&
	-2M_0^2\left(11M_0^4-7\tilde{m}^2M_0^2+2\tilde{m}^4\right)\tilde{p}^4+12M_0^4\left(M_0^2-\tilde{m}^2\right)^2\tilde{p}^2\nonumber
	\\&&+
	2M_0^2\bigg[-\tilde{p}^8+3\tilde{p}^4\left(3M_0^4-\tilde{m}^4\right)-2\tilde{p}^2\left(M_0^2-\tilde{m}^2\right)\left(7M_0^4-2\tilde{m}^2M_0^2+\tilde{m}^4\right)
	\nonumber \\&& +
	6M_0^2\left(M_0^2-\tilde{m}^2\right)^3\bigg]\log{\frac{\tilde{m}}{M_0}}+\frac{2M_0^2}{\sqrt{2M_0^2\left(\tilde{p}^2+\tilde{m}^2\right)
	-\left(\tilde{p}^2-\tilde{m}^2\right)^2-M_0^4}}\nonumber \\&&
	\bigg[6M_0^{10}-4M_0^8\left(6\tilde{m}^2+5\tilde{p}^2\right)+
	M_0^6\left(36\tilde{m}^4+38\tilde{m}^2\tilde{p}^2+23\tilde{p}^4\right)
	\nonumber \\&& -M_0^4\left(24\tilde{m}^6+18\tilde{p}^2\tilde{m}^4+11\tilde{p}^4\tilde{m}^2+9\tilde{p}^6\right)
	+M_0^2\left(6\tilde{m}^8+2\tilde{p}^2\tilde{m}^6\right.\nonumber  \\&& \left.-5\tilde{p}^4\tilde{m}^4-2\tilde{p}^6\tilde{m}^2-\tilde{p}^8\right)+
	\tilde{p}^2\left(\tilde{p}^2-\tilde{m}^2\right)^3\left(\tilde{p}^2+2\tilde{m}^2\right)\bigg]\arccos{\left(\frac{M_0^2+\tilde{m}^2-\tilde{p}^2}{2M_0 \tilde{m}}\right)}
	\Bigg\}, \nonumber \\&& 
\end{eqnarray}
\begin{eqnarray}
\Delta B^s_{h.o.}(t,m_\pi) & = & \frac{g_A^2 
	x_{\pi}^0}{96\pi^2F_{\pi}^2}\,\frac{M_N(m_{\pi})}{M_0}\,\int_{-\frac{1}{2}}^{\frac{1}{2}}\!du\,
	\Bigg\{-6M_0^2+18m_{\pi}^2\left(2\log{\frac{M_0}{\lambda}}-3\right)+t\left(11-6\log{\frac{M_0}{\lambda}}\right)\nonumber \\&&
	-\frac{36M_0^4}{\tilde{p}^8\sqrt{2\left(\tilde{m}^2+\tilde{p}^2\right)M_0^2-\left(\tilde{p}^2-\tilde{m}^2\right)^2-M_0^4}}
	\bigg[M_0^8-\left(4\tilde{m}^2+3\tilde{p}^2\right)M_0^6
	\nonumber \\&& +\left(6\tilde{m}^4+5\tilde{p}^2\tilde{m}^2+3\tilde{p}^4\right)M_0^4-
	\left(4\tilde{m}^6+\tilde{p}^2\tilde{m}^4+\tilde{p}^4\tilde{m}^2+\tilde{p}^6\right)M_0^2\nonumber \\&&+\tilde{m}^8-\tilde{m}^6\tilde{p}^2\bigg]
	\arccos{\left(\frac{M_0^2+\tilde{m}^2-\tilde{p}^2}{2M_0\tilde{m}}\right)}-\frac{6M_0^4}{\tilde{p}^8}\bigg[2\tilde{p}^6
	-3\left(3M_0^2-\tilde{m}^2\right)\tilde{p}^4\nonumber \\&&
	+6\left(M_0^2-\tilde{m}^2\right)^2\tilde{p}^2+
	6\left(M_0^2\tilde{p}^4-2M_0^2\left(M_0^2-\tilde{m}^2\right)\tilde{p}^2
	+\left(M_0^2-\tilde{m}^2\right)^3\right)\log{\frac{\tilde{m}}{M_0}}\bigg]\Bigg\},\nonumber \\&&
	+ B_{33}^r(\lambda) \,\frac{M_N(m_{\pi})}{M_0}\,\frac{4 m_{\pi}^2}{\Lambda_\chi^2} +B_{34}^r(\lambda)\, \frac{M_N(m_{\pi})}{M_0}\,
\frac{t}{\Lambda_\chi^2}, \label{db}  
\end{eqnarray}
\begin{eqnarray} 
\Delta C^s_{h.o.}(t,m_\pi) & = & \frac{g_A^2 x_{\pi}^0
	M_0^2}{96 \pi^2F_{\pi}^2}\,\frac{M_N(m_{\pi})}{M_0}\,\int_{-\frac{1}{2}}^{\frac{1}{2}}\!\frac{du}{\tilde{p}^8}\,\Bigg\{5\tilde{p}^8-9M_0^2\tilde{p}^6
	-6u^2\tilde{p}^2 \bigg[3\tilde{p}^6-2M_0^2\tilde{p}^4\nonumber
	\\ &&-3M_0^2\left(M_0^2-3\tilde{m}^2\right)\tilde{p}^2-6\left(M_0^3-M_0\tilde{m}^2\right)^2\bigg]
	 -9\left(4u^2-1\right)\tilde{p}^8\log{\frac{\tilde{m}}{M_0}}\nonumber \\ &&
	+9M_0^2\bigg[4u^2\left(-\tilde{m}^2\tilde{p}^4+2\tilde{m}^2\tilde{p}^2\left(\tilde{m}^2-M_0^2\right)
	+\left(M_0^2-\tilde{m}^2\right)^3\right)\nonumber \\&&
	-\tilde{p}^4\left(M_0^2-\tilde{m}^2\right)\bigg]\log{\frac{\tilde{m}}{M_0}}
	+\frac{9M_0^2}{\sqrt{2M_0^2\left(\tilde{m}^2+\tilde{p}^2\right)-\left(\tilde{p}^2-\tilde{m}^2\right)^2-M_0^4}}
	\bigg[4u^2M_0^8\nonumber
	\\&&-4\left(4\tilde{m}^2+\tilde{p}^2\right)u^2M_0^6-\left(-24u^2\tilde{m}^4+4u^2\tilde{p}^2\tilde{m}^2+\tilde{p}^4\right)M_0^4
	\nonumber \\&& +
	\left(-16u^2\tilde{m}^6+20u^2\tilde{p}^2\tilde{m}^4+2\left(1-2u^2\right)\tilde{m}^2\tilde{p}^4+\tilde{p}^6\right)M_0^2
	\nonumber \\&& + \tilde{m}^2\left(\tilde{m}^2-\tilde{p}^2\right)
	\left(4u^2\tilde{m}^4-8u^2\tilde{m}^2\tilde{p}^2+\left(4u^2-1\right)\tilde{p}^4\right)\bigg]\arccos{\left(\frac{M_0^2+\tilde{m}^2-\tilde{p}^2}{2M_0\tilde{m}}\right)}
	\Bigg\},\nonumber \\&&  
\end{eqnarray}
with the new variable
\begin{eqnarray}
\tilde{m}^2 & = & m_{\pi}^2+\left(u^2-\frac{1}{4}\right)t.
\end{eqnarray}
We note that the contributions of $\Delta A_{2,0}^s(t,m_\pi),\,\Delta C_{2,0}^s(t,m_\pi)$ are finite at ${\cal O}(p^3)$ in BChPT, while $\Delta B_{2,0}^s(t,m_\pi)$ contains
two new counterterms $B_{33}^r(\lambda),\,B_{34}^r(\lambda)$ at this order \cite{future}.

\vfill\pagebreak

\end{document}